%% file: main.tex
\documentclass[twocolumn]{aastex631}

\usepackage{amsmath}
\usepackage[ruled,vlined]{algorithm2e}
\usepackage{needspace}
\usepackage{etoolbox}

\newcommand{\review}[1]{#1}  

\newcounter{panel}[figure]

\makeatletter
\let\orig@sect\@sect
\let\orig@ssect\@ssect

\def\@sect#1#2#3#4#5#6[#7]#8{%
  \Needspace{2\baselineskip}%
  \orig@sect{#1}{#2}{#3}{#4}{#5}{#6}[#7]{#8}%
}

\newcommand{\KL}[2]{D_{\rm KL}(#1 \, \Vert \, #2)}

\shorttitle{NPE for Spatially Varying Backgrounds and PSFs}
\shortauthors{Patel et al.}

\begin{document}
\title{Neural Posterior Estimation for Cataloging Astronomical Images with Spatially Varying Backgrounds and Point Spread Functions}

\author[0009-0006-6476-4423]{Aakash Patel}
\affiliation{Department of Electrical Engineering and Computer Science, University of Michigan, Ann Arbor, MI 48109, USA}

\author[0000-0002-5596-198X]{Tianqing Zhang}
\affiliation{Department of Physics and Astronomy and PITT PACC, University of Pittsburgh, Pittsburgh, PA 15260, USA}

\author[0000-0001-8868-0810]{Camille Avestruz}
\affiliation{Department of Physics, University of Michigan, Ann Arbor, MI 48109, USA}
\affiliation{Leinweber Center for Theoretical Physics, University of Michigan, Ann Arbor, MI 48109, USA}

\author[0000-0002-1472-5235]{Jeffrey Regier}
\affiliation{Department of Statistics, University of Michigan, Ann Arbor, MI 48109, USA}

\author{the LSST Dark Energy Science Collaboration}
\noaffiliation


\begin{abstract}
Neural posterior estimation (NPE), a type of amortized variational inference, is a computationally efficient means of constructing probabilistic catalogs of light sources from astronomical images. 
To date, NPE has not been used to perform inference in models with spatially varying covariates. However, ground-based astronomical images exhibit spatially varying sky backgrounds and point spread functions (PSFs), and accounting for this variation is essential for constructing accurate catalogs of imaged light sources. 
In this work, we introduce a novel NPE-based cataloging method that trains an inference network with semi-synthetic astronomical images generated using PSF and backgrounds sampled from \review{from the Sloan Digital Sky Survey}.
In experiments \review{with semi-synthetic images}, we evaluate the method on key cataloging tasks: light source detection, star/galaxy separation, and flux measurement.
\review{A ``generalist'' inference network---trained with diverse PSFs and backgrounds---performs as well as a ``specialist'' network even when both are evaluated on the specialist's particular PSF/background combination. This result suggests that a single NPE network can generalize across spatial variations, eliminating the need for retraining on each observational condition.}%
\end{abstract}

\section{Introduction}  \label{sec:intro}

Astronomical imaging enables observation of distant galaxies, providing astronomers with a crucial pathway for understanding the universe. Measurements of the shape and flux of imaged galaxies are influenced by observational factors such as the point spread function (PSF) and the sky background. In ground-based astronomical images, the background and PSF both vary spatially.

The background in ground-based astronomical images varies mainly due to variation in sky glow, which arises from the scattering of light from artificial light sources \citep{popowicz2015}. Sky glow is particularly variable across the azimuth and zenith angles. Other sources of background variation include noise or internal reflection from detectors used in image capture \citep{slater2009removing}; changing atmospheric conditions, such as cloud cover coupled with light pollution; and sources beyond the atmosphere, including light from stars scattered by the PSF, galactic dust, and faint or ultra-diffuse galaxies \citep{blanton2011improved, liu2023recipe}. In addition, depending on the source, background variation can vary over different timescales; for example, atmospheric turbulence can vary over seconds, whereas light pollution varies over years. 

The PSF varies due to both changes in the observation conditions, such as atmospheric turbulence and air mass, and changes in the optical and electronic properties of the instruments used to capture images \citep{gentile2013interpolating}.
The PSF is commonly characterized in terms of its full width half maximum (FWHM), which represents the width of the PSF profile at half of its maximum value. A larger FWHM corresponds to greater blurring by the imaging system. Variation in the PSF for a ground-based telescope in terms of FWHM is often 50--100\% of its median \citep{Jarvis2021, Li2021}. Even in the absence of atmospheric effects, the FWHM of the PSF can vary significantly across a CCD, with the largest discrepancies observed in the chips farthest from the optical axis \citep{xin2018study}.

Multiple stages of the image-processing pipelines used to analyze data from modern imaging surveys require precise knowledge of the local background level and PSF, including background subtraction, object detection, star/galaxy separation, deblending (the process of attributing pixel flux to specific astronomical objects), and measurement of object photometry and shape. 
Understanding and accounting for variation in background and PSF is crucial for detecting and characterizing objects with high precision. For example, deblending with an incorrect PSF can lead to inaccurate photometry and astrometry, which can impact object detection and photometric redshift estimation. It can also impact the prediction of galaxy morphologies, which can influence weak lensing science, the study of galaxy evolution, and other applications \citep{lupton2001sdss, fischer2001weak, cypriano2010cosmic}.

Modern astronomical surveys, such as the Rubin Observatory Legacy Survey of Space and Time (LSST), require sophisticated image-processing tools to effectively characterize each observed object for two reasons. First, with the larger data volumes in modern surveys, the statistical error across multiple cosmological probes is less compared to the previous generation, leaving a smaller budget for systematic error. Second, the greater depth of modern astronomical surveys increases the number density of detectable galaxies, making deblending a more important task in the image-processing pipeline.

Bayesian methods have been shown to excel in dealing with ambiguity due to blending \citep{brewer2013probabilistic, portillo2017improved, regier2019approximate, feder2020multiband, liu2023variational, buchanan2023markov}.
However, traditional Bayesian methods typically require fitting a new posterior approximation for each image, which is computationally expensive, especially when large, complex models with many parameters are coupled with unprecedented data volume. Amortized Bayesian inference methods improve scalability by performing approximate inference more efficiently \citep{kingma2014auto}. Instead of refitting the posterior approximation for each image, amortized variational inference trains a neural network (called the ``inference network'') that can be applied over multiple images, thereby amortizing the computational cost. After training, inference on a new image requires just a single forward pass through the inference network, accelerating the image-processing pipeline.

A particular amortized variational inference technique known as neural posterior estimation (NPE) has recently gained popularity both in the fields of astronomy \citep[e.g.,][]{zhang2023nbi,lemos2023robust} and machine learning \citep[e.g.,][]{rodrigues2021hnpe}.
NPE has been shown to be effective specifically for cataloging astronomical images \citep{liu2023variational}. 
In NPE, the inference network is trained with synthetic data sampled from the model through ancestral sampling.

NPE has numerous advantages for cataloging. First, it is a likelihood-free method in that it does not require evaluation of the conditional likelihood function; instead, it suffices to draw samples from the joint distribution. This means even complex simulators, such as GalSim \citep{galsim2015} and PhoSim \citep{peterson2015simulation}, can be used to encode the conditional likelihood. Second, NPE enjoys important theoretical guarantees that are not shared by traditional (ELBO-based) approaches to amortized variational inference, which minimize the reverse Kullback-Leibler (KL) divergence, that is, the KL divergence from the posterior approximation to the posterior. NPE instead minimizes the forward Kullback-Leibler divergence and marginalizes over nuisance latent variables\footnote{
Nuisance latent variables are unobserved model parameters that are necessary for accurate forward modeling but are not themselves of direct scientific interest. In our work, the PSF and background intensity levels are nuisance variables: they reflect properties of the instrumentation and observing conditions rather than of the astronomical objects we aim to catalog.
}
implicitly by simply excluding them from the objective function \citep{ambrogioni2019forward}. Further, with reasonable conditions, NPE is globally convergent, whereas other types of variational inference often converge to shallow local optima \citep{mcnamara2024globally}.

To date, NPE has not been used to catalog astronomical images with spatially varying covariates, such as backgrounds and PSFs, despite the importance of doing so. In this work, we propose a statistical model that treats the background and PSF of images, along with the latent properties of imaged light sources, as random variables in a Bayesian model (Section~\ref{sec:stat_model}). In this model, the backgrounds and PSFs follow the empirical distribution of backgrounds and PSFs estimated by an existing survey.
\review{In principle, background samples could come from any source, including both the prior and posterior of sophisticated structured models of \citet{saydjari2022photometry} and \citet{feder2023pcat}. Here, however, we focus on the empirical distribution of background estimates produced by the Sloan Digital Sky Survey's photometric pipeline, which exhibits more gradual spatial variation than samples from these structured models.
}

Additionally, we propose an NPE procedure to train an inference network to estimate the parameters used to generate the data, such as the number of objects in the image and their positions and fluxes (Section~\ref{sec:inference}).
One challenge lies in designing an inference network capable of extracting features on two vastly different spatial scales. Light sources usually cover hundreds of pixels (e.g., a 20$\times$20-pixel patch), while point spread functions (PSFs) and backgrounds vary slowly and exhibit spatial dependencies that span millions of pixels and multiple images.
To accurately infer the properties of a light source, an inference network must attend to the broader context, to learn about the background and PSF, and to the highly localized image region that pertains to the particular light sources.
A sophisticated network design would be needed to operate on such different scales.
If multiple images are required to infer the broader context, a sophisticated data loading scheme would be needed as well.

We circumvent the need for such sophistication by combining traditional background and PSF estimation routines with an off-the-shelf CNN architecture. Specifically, we provide our inference network with estimates of the background and PSF determined by existing routines along with the image data. The network can then use or ignore the background and PSF inputs as it sees fit. Since the images to be cataloged are small, we anticipate that these globally derived inputs will be useful side information.

Once trained, this inference network takes images, survey-estimated backgrounds, and survey-estimated PSFs as input. It outputs a posterior approximation in a single forward pass, without any additional training.
We demonstrate the effectiveness of our method by applying it to \review{semi-synthetic} images designed to \review{mimic images from} the Sloan Digital Sky Survey (SDSS) Section~\ref{sec:experiments}). Finally, we discuss the limitations of our work and propose extensions to improve the proposed method (Section~\ref{sec:discussion}).

\section{The SDSS Image-Processing Pipeline}  \label{sec:sdss_psf}
The semi-synthetic data that we generate for training and validating our method is based on data from the Sloan Digital Sky Survey (SDSS). The SDSS is a spectroscopic and astronomical imaging survey designed to study the redshift of extragalactic content.
SDSS data have been used for a variety of research objectives, including investigations of galaxy formation and evolution, quasars, and the evolution of dark energy. While our method is adaptable to data from any survey, the SDSS offers an ideal test case due to its well-characterized properties and long operational history.

\review{
In SDSS, photometric completeness for galaxies decreases notably around 22 r-band magnitude. In DR7, S11 field, comparisons with the deeper COMBO‑17 survey indicate that SDSS recovers only about 50\% of galaxies at 22 r-band magnitude \citep{sdss2006completeness}. Further, \citet{johnston2007completeness} introduced formal magnitude‑completeness tests, which statistically validate that SDSS reaches its claimed faint limits. Although SDSS attains a $5\sigma$ point-source sensitivity of 22.7 r-band magnitude, this pertains solely to isolated stars; extended galaxy completeness declines significantly by 22 magnitude, marking the onset of a regime characterized by lower signal-to-noise and reduced detection efficiency.
}

\review{
In SDSS, star/galaxy separation performance has been benchmarked in multiple ways. The original Photo pipeline classified sources based on the difference between PSF and model magnitudes, aiming to distinguish point-like and extended sources \citep{lupton2001sdss}. This morphological separation is reliable down to $r \sim 21$, beyond which classification accuracy diminishes due to increasing similarity between compact galaxies and stars \citep{scranton2002analysis, yahata2007effect}. \citet{ball2006robust} introduced a machine learning approach using decision trees trained on a spectroscopic sample from SDSS DR3. Their classifier achieved galaxy completeness and efficiency of 98.9\% and 98.3\%, respectively, and star completeness and efficiency of 91.6\% and 93.5\%, respectively, within the magnitude range of $r < 17.77$. They reported consistent performance to $r \sim 20$, with decreasing reliability at fainter magnitudes.
}

In addition to these benchmarks, the SDSS pipeline's estimates of background intensity and PSF are important for our work.

\textbf{Background estimation.}
Beginning with SDSS Data Release 8 in January 2011, the method introduced by \citet{blanton2011improved} has been used to estimate the background. This method has two steps. First, bright stars and galaxies are identified and masked. Objects brighter than magnitude 15 are found using the SDSS catalog, and masks are created with larger sizes for brighter objects (which contribute a non-negligible amount of flux to more pixels). An external catalog is used to identify and mask additional large galaxies that might otherwise affect background measurements. Further, to address light reflecting off the edges of SDSS filters, stars just outside the frame are identified using the Tycho-2 catalog, and the affected areas are masked. For exceptionally bright objects that disrupt processing by the SDSS photometric pipeline, entire fields are masked.

In the second step, the data is binned into larger pixel blocks, and a spline model is fitted to them.
Regions that significantly deviate from the model fit are further masked, and the fitting process is iterated multiple times.

\textbf{PSF estimation.}
The SDSS pipeline, known as \texttt{photo}, models PSFs using the Karhunen-Lo\'eve decomposition to decompose stars brighter than 20 magnitude into eigenimages that retain the first three components \citep{lupton2005sdss}. \review{The spatial variation of a PSF across an image is captured by modeling the coefficients of these components as low-order polynomials in position. From this spatially varying, pixelized PSF model,} the pipeline then fits a smooth, radially symmetric mixture model composed of two Gaussians and a power-law tail, \review{with six parameters in total: $\sigma_1$ and $\sigma_2$ are standard deviations for the two Gaussians, $b$ is the ratio of the second Gaussian to the first at the origin, $\sigma_p$ is the width parameter for the power-law model, $p_0$ is the value of the power law at the origin, and $\beta$ is the slope of the power law.}
At radius $r$, this PSF model is
\begin{align*}
    \mathbf{\Pi}(r) &= \frac{\exp{\left(-\frac{r^2}{2\sigma_1^2}\right)} + b\exp{\left(-\frac{r^2}{2\sigma_2^2}\right)}
  + p_0 \left(1 + \frac{r^2}{\beta\sigma_p^2}\right)^{-\frac{\beta}{2}}}{1 + b + p_0}.
\end{align*}

\begin{figure}
    \centering
    \includegraphics[width=\linewidth]{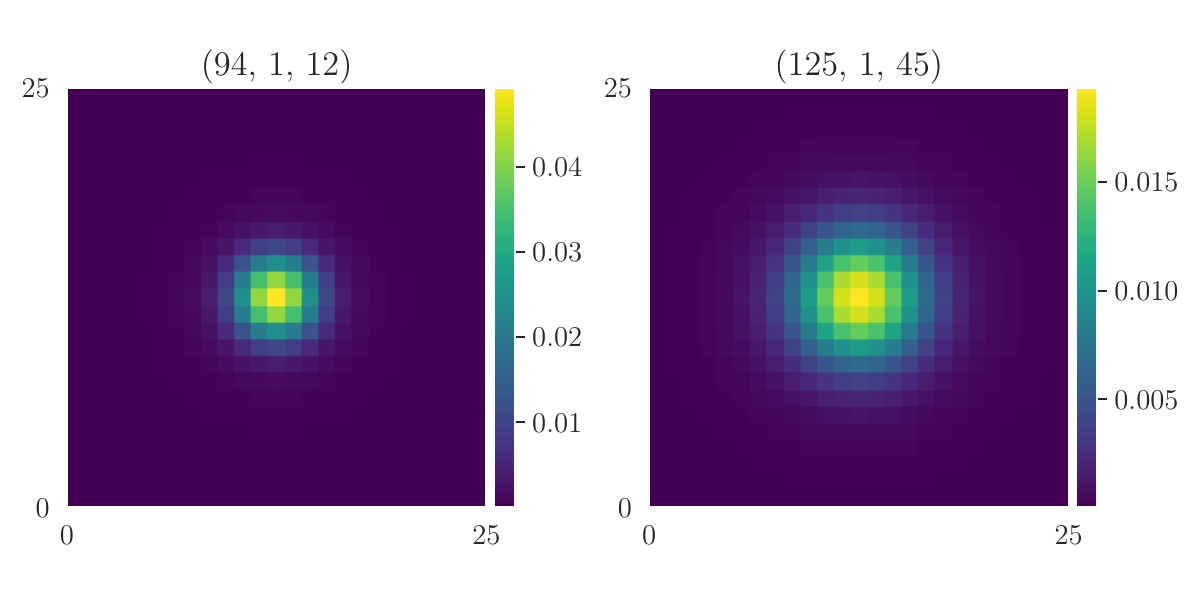}
    \caption{\textbf{Pixelated PSFs for two distinct SDSS fields.} The run number, camera column, and field number, which together identify a particular SDSS field, appear above each image. PSF variation is due to variation in the observing conditions and the telescope optics.}
    \label{fig:psfs}
\end{figure}

\begin{figure}
    \centering
    \includegraphics[width=\linewidth]{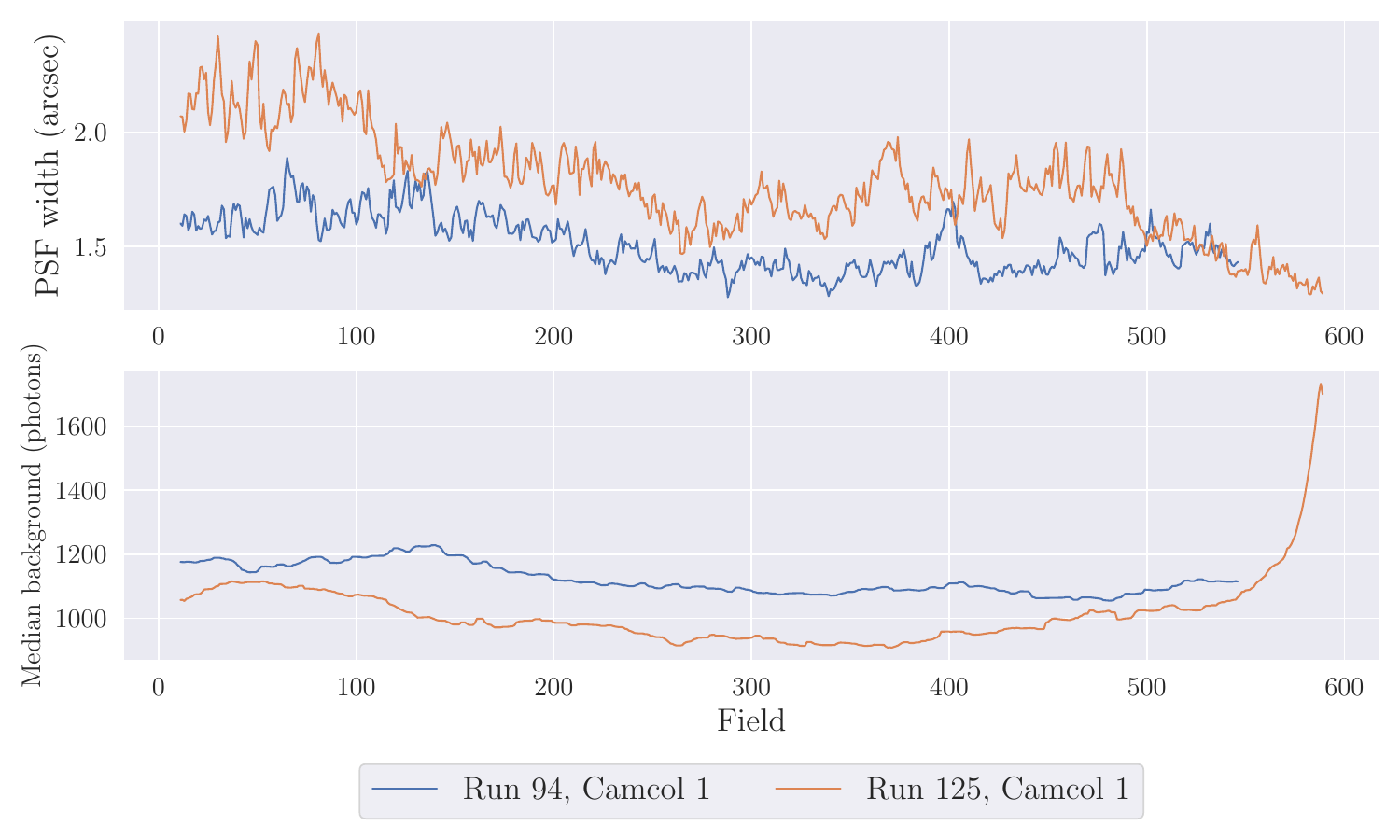}
    \caption{The SDSS image data collected during a particular nightly run, for each camera column (camcol), is segmented temporally into fields. The PSF width and background intensity vary significantly within and across these runs.
    Plots like these for PSF width appeared previously in \citet[Figure 8]{earlySDSS2002}.}
    \label{fig:psf_bg_variation}
\end{figure}

Figure~\ref{fig:psfs} shows PSFs for two SDSS fields. There is a visible difference in their sizes. The background and PSF variation for multiple nightly runs of the Sloan Digital Sky Survey (SDSS) are illustrated in Figure~\ref{fig:psf_bg_variation}. The wide variation in both the PSF size and background level in the SDSS dataset further illustrates its suitability as a dataset for validation of our method.

\section{Methodology}  \label{sec:methodology}

\begin{figure*}
    \centering
    \includegraphics[width=\linewidth]{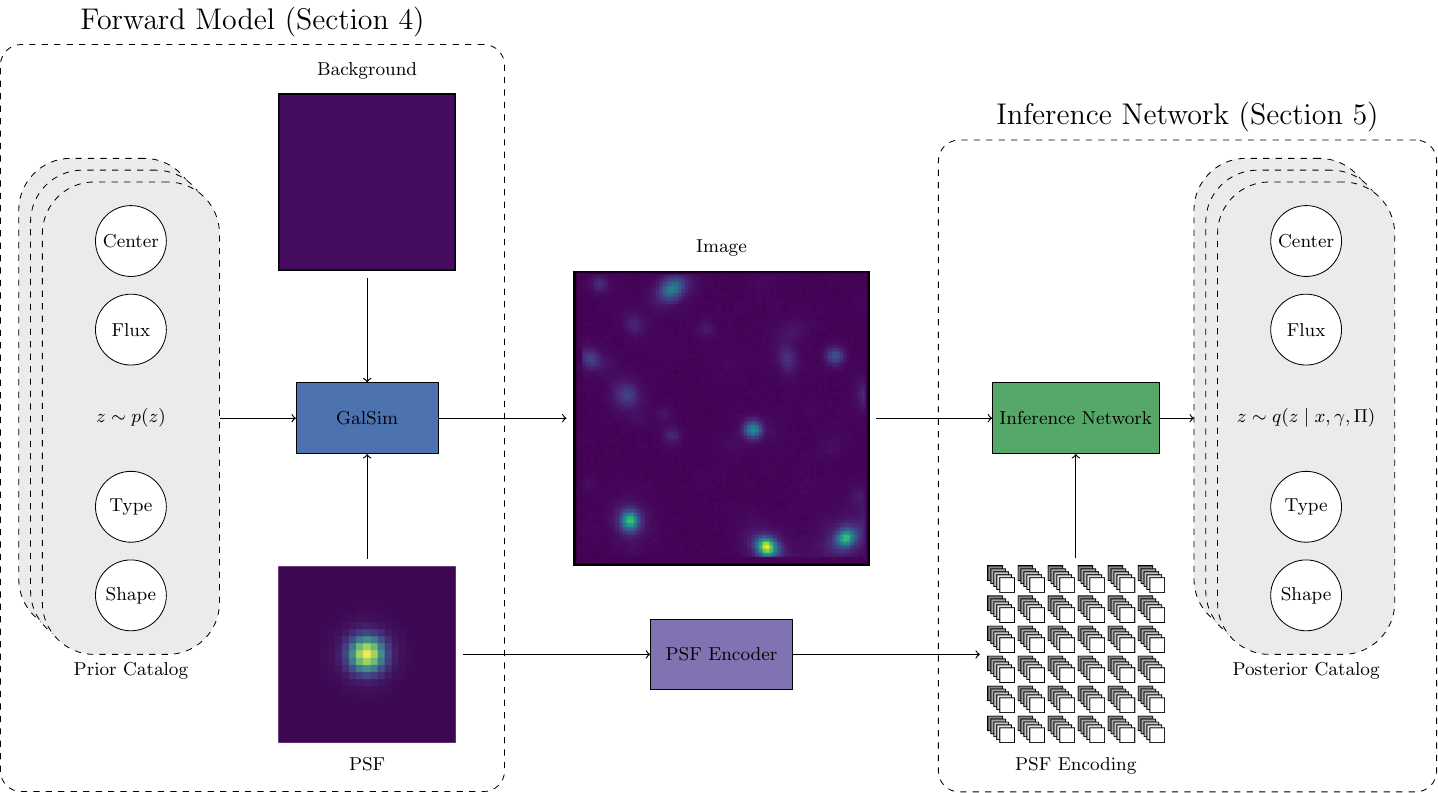}
    \caption{\textbf{Overview of the inference procedure.} The catalog (light grey, left) follows a parametric prior, whereas the background and PSF follow SDSS empirical distributions; they are sampled by selecting an SDSS field uniformly at random. Together, a catalog, PSF, and background define a sky image that is generated by GalSim, which adds Poisson shot noise. The inputs to the inference network are the image and the PSF. The inference network outputs a posterior approximation of the catalog conditioned on the image and PSF (light gray, right).}
    \label{fig:end_to_end_model}
\end{figure*}

Our method has two components: a forward model for simulating synthetic astronomical images and an inference network for inferring a posterior catalog from those images (Figure~\ref{fig:end_to_end_model}). The two components work in tandem: the forward model is used to generate realistic training data with spatially varying backgrounds and PSFs, while the inference network inverts this process, inferring catalogs from the simulated images.

In the forward model, catalogs, backgrounds, and PSFs follow a given prior distribution. The conditional likelihood is encoded implicitly by GalSim \citep{rowe2015galsim}, which generates synthetic images given the sampled catalogs, backgrounds, and PSFs. The parameters of the forward model are not learned or updated during training; the forward model is only sampled to produce paired catalogs, backgrounds, PSFs, and synthetic astronomical images. The forward model is described in more detail in Section~\ref{sec:stat_model}.

The inference network is trained using this simulated data. The network takes the image, background, and PSF as inputs and outputs an approximation of the posterior distribution across catalogs. In contrast to the forward model, the inference network is trainable, with parameters optimized by stochastic gradient descent. The network is trained to minimize the negative log-likelihood (NLL) of the sampled catalog under the posterior approximation. The inference network is described in more detail in Section~\ref{sec:inference}.

\section{Forward Model: Training Data Generation}  \label{sec:stat_model}
Most probabilistic cataloging approaches employ similar forward models \citep{brewer2013probabilistic, lang2016tractor, regier2019approximate, portillo2017improved, feder2020multiband, liu2023variational, buchanan2023markov}. Our model specifically extends the model of \citet{hansen2022scalable}, which does not treat background and PSF as random.

\subsection{Prior Distribution}  \label{sec:prior}
We propose a prior distribution over catalog $z$, background intensity $\gamma$, and PSF $\Pi$ with the following factorization:
\begin{align*}
    p(z, \gamma, \Pi) := p(z)p(\gamma, \Pi).
\end{align*}
In other words, the catalog, which includes the latent properties of each light source, is sampled independently of the field-level properties (i.e., the background and PSF).

Our catalog prior has the form
\begin{align}
    p(z) := p(|z|)\prod_{s = 1}^{|z|} p(z_s),
\end{align}
where $|z|$ denotes the cardinality of set $z$, i.e., the number of imaged sources. The distribution of $|z|$ is
\begin{align}
    |z| \sim \mathrm{Poisson}(\mu HW),
\end{align}
where $H$ and $W$ are the height and width of the image in pixels and $\mu = 6.25 \times 10^{-4}$, to resemble the typical per-pixel source density in SDSS.
\review{For simplicity, $\mu$ is fixed in our work; modeling $\mu$ as random could better represent spatial variation in source density.}
For each $s=1,\ldots,|z|$, the properties of light source $s$, denoted $z_s$, include position, source type (star or galaxy), flux, and galaxy shape. They are sampled according to the specification in Table~\ref{tab:prior-factors}, which is fitted to earlier catalogs of SDSS.

We set $p(\gamma, \Pi)$ to the empirical joint distribution of backgrounds and PSFs in a diverse subset of SDSS fields.
Sampling a background and PSF entails choosing an SDSS field uniformly at random and then returning the background and PSF from the \texttt{photo} pipeline for that field.
\review{
An alternative choice would be to set $p(\gamma, \Pi) := p(\gamma) p(\Pi)$, that is, the product of marginals, and to sample backgrounds and PSFs independently from SDSS.
This alternative prior ignores potential correlation between background and PSF, and thus produce a overdispersed prior. Overdispersed priors are widely used in statistics, and ignoring correlation in the prior does prevent recovering correlation in the posterior. However, because we have a ample supply of paired backgrounds and PSFs, we do not explore this-data efficient alternative.
}

For simplicity, for each SDSS field, we consider only the modeled PSF for the field center, rather than selecting the PSF for a random position within each field.

\begin{table*}
    \centering
    \begin{tabular}{@{}l|l|l|l@{}}
Latent variable & Distribution & Parameter(s) & Units\\
\noalign{\hrule height 0.04em}
Position & Uniform & support: [0, 1) $\times$ [0, 1) & - \\
Is a galaxy & Bernoulli & rate: 0.514 & -\\
Star flux & Truncated Pareto & b: 0.469; c: 1000: loc: -0.553; scale: 1.185 & nmgy\\
Galaxy flux & Truncated Pareto & b: 1.561; c: 1000; loc: -3.29; scale: 3.29 & nmgy\\
Galaxy angle & Uniform & support: [0, pi) & radians \\
Galaxy disk fraction & Uniform & support: [0,1] & - \\
Disk axis ratio & Uniform & support: (0, 1) & - \\
Disk half-light radius & Gamma & shape: 0.393; loc: 0.837; scale: 4.432 & arcsec \\
Bulge axis ratio & Uniform & support: (0, 1) & - \\
Bulge half-light radius & Gamma & shape: 0.393; loc: 0.419; scale: 2.216 & arcsec \\
    \end{tabular}
\caption{
The prior distribution of light source properties. These properties are sampled independently for each light sources, independently of the field-level properties (i.e., background and PSF).
}
\label{tab:prior-factors}
\end{table*}

\subsection{Conditional Likelihood}  \label{sec:likelihood}
Astronomical images record the \review{the intensity of electromagnetic radiation} in a given pixel. The intensity of each pixel depends on three factors: the properties of the light sources in the image (e.g., positions, source types, fluxes, and shapes), the background intensity, and the point spread function. 

Given a catalog $z$, background $\gamma$, and PSF $\Pi$, we model the \review{count of photoelectrons} $x_{ij}$ at pixel $(i,j)$ as a Poisson process with a rate $\lambda_{ij}$ that depends on $z$, $\gamma$, and $\Pi$:
\begin{align*}
    x_{ij} \mid z, \gamma, \Pi \sim \text{Poisson}(\lambda_{ij})
\end{align*}
Specifically, we define
\begin{align}
    \lambda_{ij} := \gamma_{ij} + \sum_{s=1}^S \xi_{ij}(z_s, \Pi). \label{eq:calc_rate}
\end{align}
where $\gamma_{ij}$ is the background intensity for pixel $(i,j)$ and $\xi_{ij}$ is a deterministic function that maps a source's properties to its expected contribution to pixel $(i,j)$ based on the PSF. This deterministic function is encoded by the GalSim image simulator \citep{rowe2015galsim}. Since \review{the background intensity $\gamma_{ij}$ is always at least several hundred}, we use a Gaussian approximation to the Poisson distribution:
\begin{align}
    x_{ij}' \sim \mathcal{N}(\lambda_{ij}, \lambda_{ij}) \label{eq:sample_pixel}
\end{align}

Finally, we subtract the background from the image:
\begin{align*}
    x_{ij} := x_{ij}' - \gamma_{ij}
\end{align*}
Note that this is not the same as simulating images with no background. Since the variance at a pixel also depends on the background intensity at that pixel, sky-subtracted images whose (subtracted) background has greater intensity will still have greater variance than sky-subtracted images with lower background intensities.

The forward model is illustrated in Figure~\ref{fig:end_to_end_model}. A procedure for generating our simulated dataset is given by Algorithm~\ref{alg:sampling}.
Because our approach to inference (described in the next section) is likelihood-free, this conditional likelihood is never explicitly evaluated; it is only used implicitly in our procedure for generating synthetic data.

\begin{algorithm}
\caption{Sampling for data simulation} \label{alg:sampling}
\KwIn{Number of samples $N \in \mathbb{N}$, SDSS field $K \in \mathbb{N}$}
Initialize simulated dataset $D \gets \varnothing$\\
\For{$n = 1, \dots, N$}{
Sample catalog $z \sim p(z)$\\
Choose SDSS field $k \sim \mathrm{Uniform}(1, \dots, K)$\\
$\Pi \gets$ \texttt{photo} PSF for field $k$\\
$\gamma \gets$ \texttt{photo} background for field $k$\\
\For{\textnormal{pixel row} $i = 1, \ldots, H$}{
\For{\textnormal{pixel column} $j = 1, \ldots, W$}{
Calculate rate $\lambda_{ij}$ according to Equation~\ref{eq:calc_rate}\\
Sample pixel $x_{ij}$ according to Equation~\ref{eq:sample_pixel}\\
Perform sky subtraction: $x_{ij} := x_{ij} - \gamma_{ij}$
}
}
$D \gets D \cup \{ (x, \Pi, \gamma) \}$
}
\Return $D$
\end{algorithm}

\section{Neural Posterior Estimation: Bayesian Inference with a Neural Network} \label{sec:inference}
Given a sky-subtracted astronomical image $x$ and PSF $\Pi$, we aim to infer the posterior distribution of the catalog, $p(z \mid x, \Pi)$.
This posterior is intractable as it involves integrating over all possible catalogs. Therefore, instead of inferring the posterior exactly, we approximate the posterior using neural posterior estimation (NPE). NPE aims to select a variational approximation $q_\phi$ indexed by parameters $\phi \in \Phi$ that minimizes an expectation of the forward Kullback-Leibler (KL) divergence:
\begin{align*}
    &\mathbb E_{(x, \pi, z) \sim p(x, \pi, z)} \KL{p(z \mid x, \pi)}{q_\phi(z \mid x, \pi)}\\
    &= \mathbb E_{(x, \pi, z) \sim p(x, \pi, z)} \log q_\phi(z \mid x, \pi) + c,
\end{align*}
where $c$ is constant with respect to $\phi$.

We accomplish this by training a neural network to map $(x, \Pi)$ to a parameterized representation of $q(z \mid x, \Pi)$, which we describe further in Section~\ref{sec:variational-family}.
This network, whose architecture is described in Section~\ref{sec:architecture}, is trained through stochastic optimization using the simulated data described in Section~\ref{sec:stat_model}.
After the network has been trained, evaluating the posterior for a given data point $(x, \Pi)$ just involves a single forward pass.

\subsection{The Variational Family}
\label{sec:variational-family}
The probabilistic structure of our variational distribution, which factorizes over small spatial regions called ``tiles,'' matches that of \cite{liu2023variational}.
In particular, let $H' = \lceil H / 4 \rceil$ and $W' = \lceil W / 4 \rceil$.
For $h=1,\ldots,H'$ and $w=1,\ldots,W'$, let image tile $T_{h,w} := [h,h+4) \times [w,w+4)$ refer to a particular 4$\times$4-pixel region. Collectively, the image tiles are a partition of the region imaged by $x$, expressed in pixel coordinates.
For a catalog $z$, let $z_{h,w} \subset z$ denote the cataloged light sources whose centroids are contained in $T_{h, w}$.

To approximate $p(z \mid x, \Pi)$, we use a family of variational distributions that factorize over tiles. In particular, we set
\begin{align*}
    q(z \mid x, \Pi) := \prod_{h = 1}^{H'} \prod_{w=1}^{W'} q(z_{h,w} \mid x, \Pi),
\end{align*}
where the probabilistic structure of $q(z_{h,w} \mid x, \Pi)$, which we refer to as the per-tile variational distribution, is given in Table~\ref{tab:dist_factors}. For each tile, this distribution is governed by eight distribution parameters. For a tile with index $(h, w)$, these distributional parameters are outputted in spatial position $(h, w)$ by a fully convolutional inference network that takes $x$ as input. We describe its architecture next.

One limitation of this tile-based variational distribution, acknowledged in \cite{liu2023variational}, is that for a light source positioned directly on a boundary between tiles, it can be ambiguous which tiles contain its centroid; this can be problematic because the detections in neighboring tiles are independent in the variational distribution. 
In ongoing work, we are addressing this limitation by developing a variational distribution that incorporates dependencies among neighboring tiles. However, for this study, we adopt this simpler form of the variational distribution, as our focus here is on modeling spatially varying covariates.

\begin{table}
    \centering
    \begin{tabular}{@{}l|l|l@{}}
        Latent variable & Distribution & \# of Parameters \\
        \noalign{\hrule height 0.04em}
        Has a source & Bernoulli & 1 \\
        Position & Trunc. Bivariate Normal & 4\\
        Is a galaxy & Bernoulli & 1\\
        Flux & Log Normal & 2\\
    \end{tabular}
    \caption{Factors of the per-tile variational distribution.}
    \label{tab:dist_factors}
\end{table}

\subsection{Neural Network Architecture}\label{sec:architecture}
Our inference network is a fully convolutional neural network whose architecture is adapted from the popular YOLO network \citep{redmon2016you}.
Although YOLO does not perform Bayesian inference, it does solve a per-tile detection problem that resembles ours, and it has been highly optimized, so we take it as a starting point.

We adapted the YOLO v5 architecture to our inference task through both systematic grid searches and less-systematic trial-and-error, in each case monitoring runtime and validation-set performance. Ultimately, we arrived at an architecture consisting of a sequence of 48 convolutional blocks, each comprising a convolutional layer, batch normalization, and SiLU activation. These blocks are followed by a final convolutional layer that outputs the predicted parameters of the posterior distribution.

The input to our inference neural network is a sky-subtracted $80 \times 80$-pixel image (i.e., $H = 80$ and $W = 80$) and a PSF (see Figure~\ref{fig:end_to_end_model}).
\review{
We use a fully convolutional inference network to ensure strong translation equivariance, so sources produce consistent outputs regardless of where they appear.
A straightforward way to incorporate the PSF would be to pad a rasterized version of it to match the input image resolution, then concatenate it to the image as an extra channel. 
This is the approach taken by \citet{lanusse2021deep} in estimating galaxy morphologies from postage stamp images.
However, estimating galaxy morphologies involves outputting a single prediction for the entire input image, whereas our network is responsible for predicting values for many sources appearing in different positions.
For our problem, if we simply concatenate a rasterized PSF to the image, the network must integrate local information about individual source from the image channel while simultaneously needing long-range context from the PSF channel. Standard CNNs are optimized for localized processing and are not well-suited to integrate such long-range dependencies.
}

\review{
Instead, we provide the PSF to the inference network in its six-parameter form, which the SDSS pipeline fits to rasterized PSF images (see Section~\ref{sec:sdss_psf}). Although the rasterized PSF generated the images, we now use the six‑parameter representation for inference. In Figure~\ref{fig:end_to_end_model}, we visualize this six-parameter PSF representation as produced by a generic PSF encoder, but at present, for simplicity, we take the PSF encoder to be the SDSS pipeline's procedure for fitting six-parameter PSF models to rasterized PSFs.
}
We discuss the possibility of using an auxiliary autoencoder as a more sophisticated method of obtaining this encoding in Section~\ref{sec:discussion}.

\review{
The two inputs---an $80\times{}80$-pixel input image and the length-6 PSF parameter vector---have different structures, so they cannot simply be stacked and provided as different channels to a CNN.
Instead, we provide each of these six PSF parameters to the CNN with a dedicated feature map (channel) that is concatenated with the image along the channel axis.
Each feature maps consists of the same value at each position and has the same height and width as the image. With this formulation, the PSF information can be accessed in the same way by a CNN at each pixel position.
}

\section{Numerical Experiments}  \label{sec:experiments}

\review{Through numerical experiments, we aim to assess the tradeoffs between training inference networks on narrowly focused versus diverse data distributions. Specifically, we compare networks trained on a single PSF and background---specialized to a particular observational setting---to those trained on a broad range of PSFs and backgrounds. Our main question is whether a generalist network, trained across heterogeneous conditions, can match the performance of a specialist network when evaluated on the specialist’s own domain. If so, this would obviate the need to train bespoke networks for each observed PSF--background combination.}



We assess the performance of various inference network training regimes in terms of precision and recall in detection (Section~\ref{sec:detection_results}), star/galaxy separation accuracy (Section~\ref{sec:classification_results}), and flux measurement error (Section~\ref{sec:flux_results}).

To generate synthetic datasets for training and validation, we sampled two variants of our prior distribution.
Both variants adhere to the description in Section~\ref{sec:prior} but differ in the collection of SDSS fields they draw on to define the empirical distributions of backgrounds and PSFs.
One prior, called $p_{\text{one}}$, samples the background and PSF from a single SDSS field (run 94, camcol 1, field 12); it serves as a baseline.
The other prior, called $p_\text{many}$, samples from $500$ SDSS fields.
As can be seen in Figure~\ref{fig:psf_bg_variation}, the field $p_{\text{one}}$ is based on is fairly typical in terms of PSF width or background level: $p_{\text{many}}$ has support for fields with wider and narrower PSFs and brighter and fainter background levels.
For simplicity, we restrict our data to the \textit{r} band in SDSS, which corresponds to a wavelength centered at 6166 angstroms \citep{earlySDSS2002}.

\begin{table*}
    \centering
    \begin{tabular}{@{}l|l|l@{}}
        Network name & Background and PSF for training & PSF availability for inference \\
        \noalign{\hrule height 0.04em}
        Single-field & a single SDSS field & not provided\\
        PSF-unaware & many SDSS fields &  not provided \\
        PSF-aware & many SDSS fields & encoding provided \\
    \end{tabular}
    \caption{Summary of the inference networks used for the numerical experiments.}
    \label{tab:networks}
\end{table*}

Using data generated from these two priors, we train three inference networks, which are summarized in Table~\ref{tab:networks}. The ``single-field'' inference network is trained with simulated images sampled from $p_{\text{one}}$, while the ``PSF-unaware'' and ``PSF-aware'' inference networks are trained with simulated images sampled from $p_{\text{many}}$.
The PSF-unaware network is not provided with image-specific PSF information for the inputted images during training and inference, whereas the PSF-aware network is; it takes image-specific PSF information as additional input.

Each network is trained with 210k simulated images for 50 epochs using the Adam optimizer and an initial learning rate of $1 \times 10^{-3}$, which decays by a factor of 0.1 after 32 epochs. Early stopping regularization is based on a 30k image held-out validation image set.
Training times are similar for these three networks, with each network requiring roughly 10 hours to train using a single NVIDIA 2080 Ti GPU.

\review{
Our primary question is whether either of the generalist networks (PSF-unaware and PSF-aware) can keep up with the specialist network (single-field) in the setting in which the specialist network specializes: with held-out data sampled from the same population as its training data was. A secondary question is whether providing a generalist network with explicit PSF information as input is necessary: it seems like it would be, but it is possible that PSF information could be learned implicitly from the input image, or that downstream metrics are insensitive to PSF variation. A final question is whether training networks with a variety of PSFs and backgrounds is even necessary, or whether a network trained with a particular PSF and background can perform well out-of-distribution and beyond the support of its training data (that is, on PSFs and backgrounds unlike those it encountered during training). We consider this third question be worthy of verification, but of secondary importance because a priori there is strong evidence that its answer in negatory: neural networks are widely known to perform poorly in the presence of domain shift \citep{zhou2022domain}.
}

\subsection{Diverse Backgrounds and PSFs}
\label{sec:point-estimates}

We first compare our three inference networks on synthetic data sampled from the $p_{\text{many}}$ prior.
We match centroids of detected sources to ground-truth sources by using the \review{Hungarian algorithm \citep{kuhn1955hungarian}} to find a bipartite matching that minimizes the sum of the distances.\footnote{
We do not take a position on whether finding an optimal bipartite match is important. Because the Hungarian algorithm runs efficiently in our setting and is readily available via the scipy Python package, we do not explore inexact alternatives. To ensure fair comparisons, the same matching algorithm is used for all evaluations.
}

\subsubsection{Light Source Detection} \label{sec:detection_results}

\begin{figure*}
\centering
\begin{minipage}{0.49\linewidth}
    \centering
    \includegraphics[width=\linewidth]{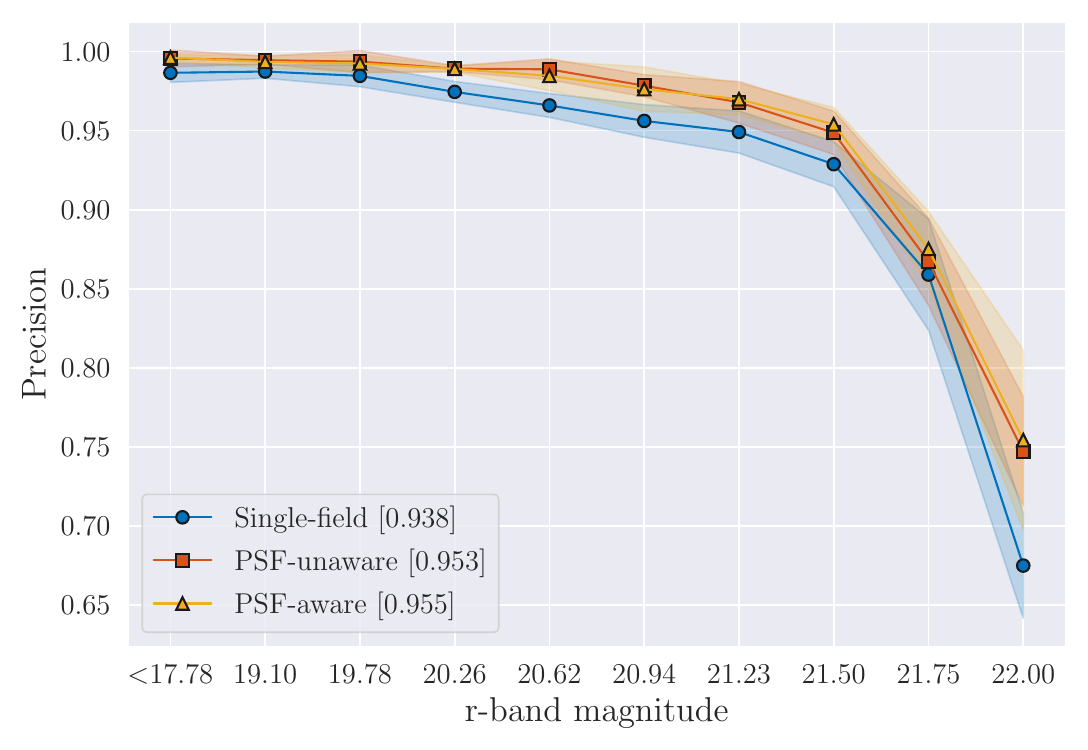}
    \par\vspace{0.5ex}
    (a) Precision for light source detection.
\end{minipage}
\hfill
\begin{minipage}{0.49\linewidth}
    \centering
    \includegraphics[width=\linewidth]{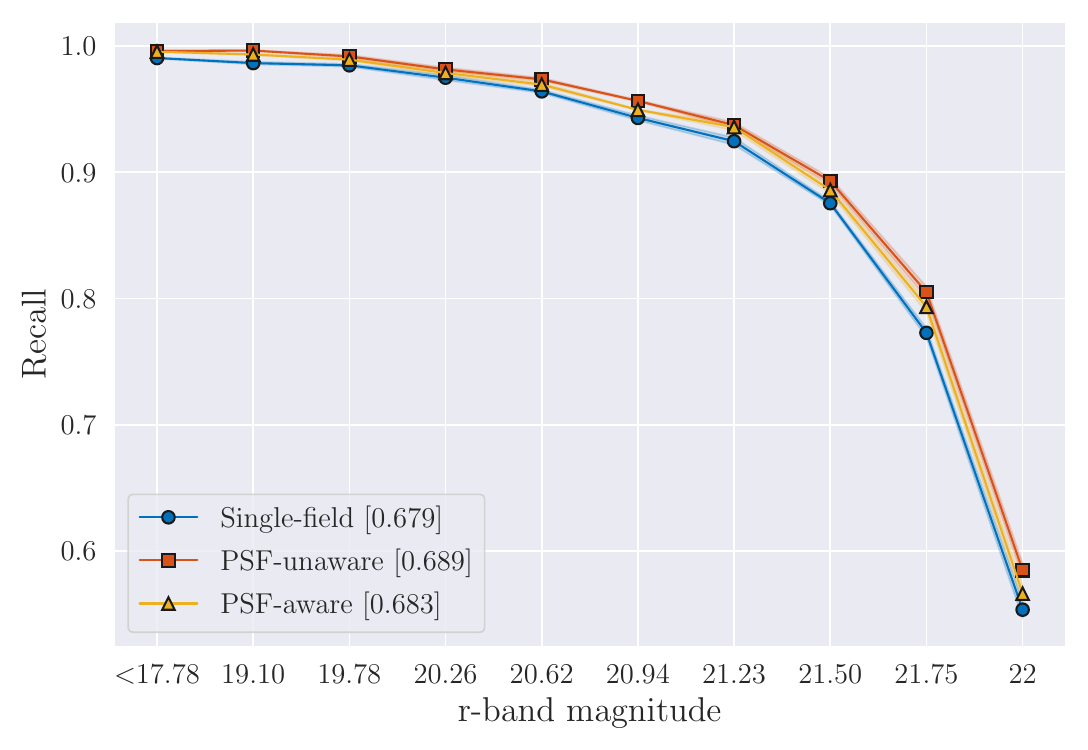}
    \par\vspace{0.5ex}
    (b) Recall for light source detection.
\end{minipage}

\vspace{1em}

\begin{minipage}{0.49\linewidth}
    \centering
    \includegraphics[width=\linewidth]{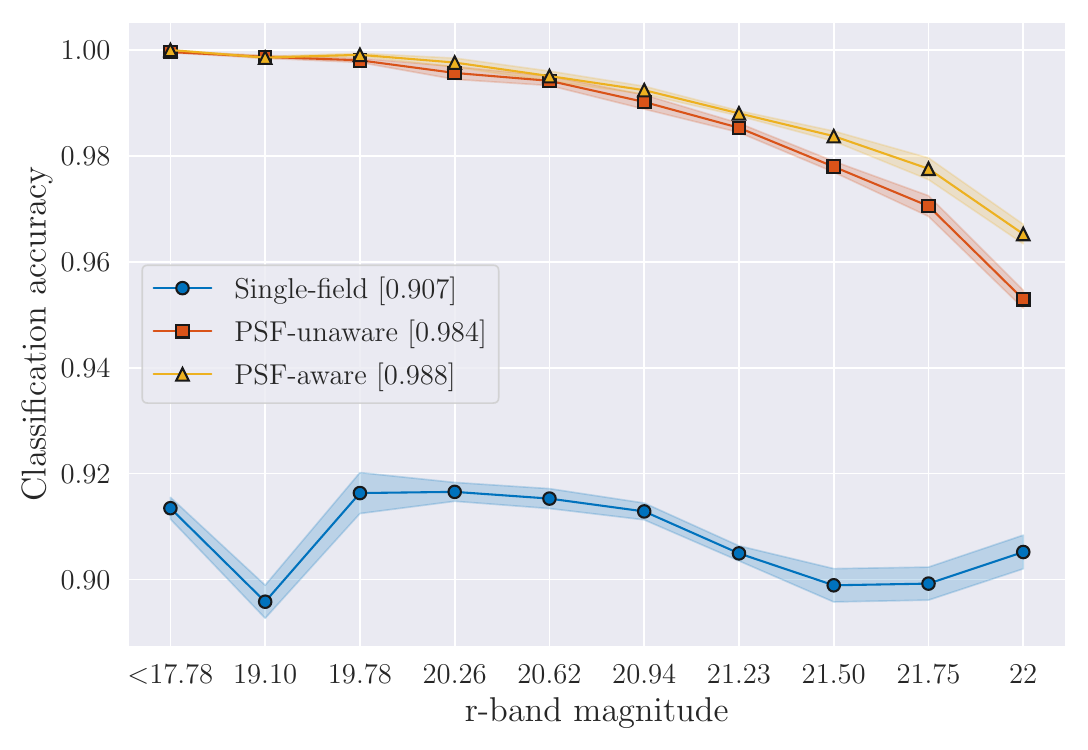}
    \par\vspace{0.5ex}
    (c) Accuracy for star/galaxy separation.
\end{minipage}
\hfill
\begin{minipage}{0.49\linewidth}
    \centering
    \includegraphics[width=\linewidth]{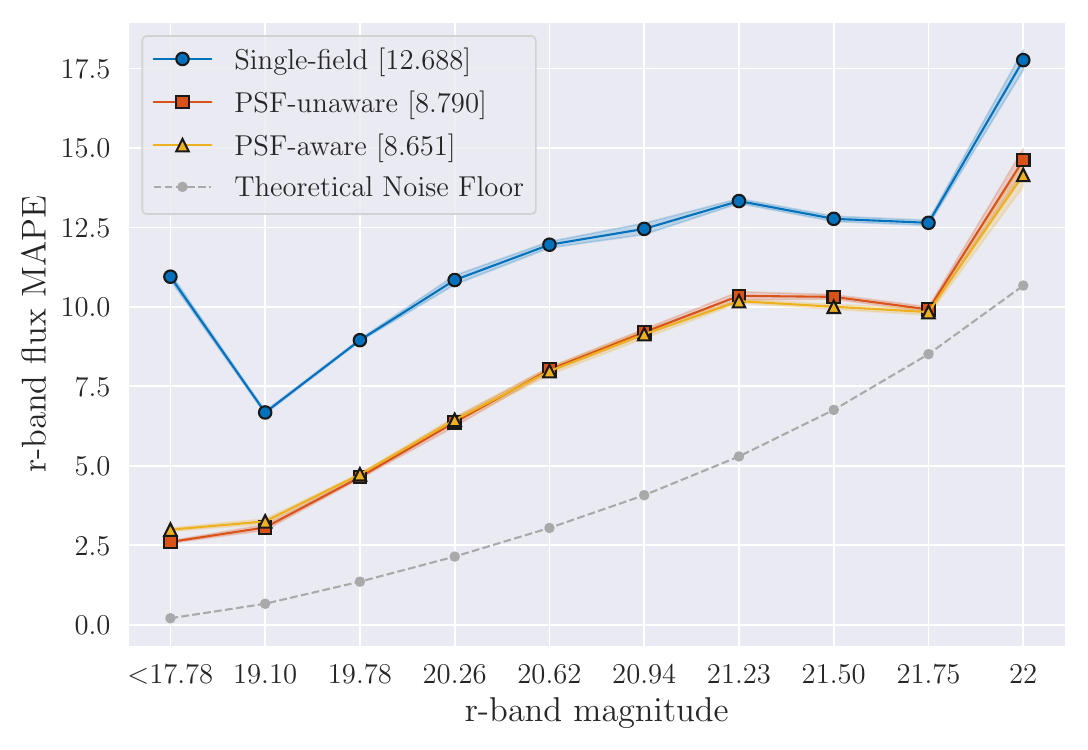}
    \par\vspace{0.5ex}
    (d) Mean absolute percentage error for flux estimation.
\end{minipage}

\vspace{2ex}

\caption{\textbf{Performance for held-out data with varied backgrounds and PSFs.}
Solid lines indicate the mean performance in each r-band magnitude bin. The widths of the bins are selected so that each bin contains an equal number of objects. The shaded regions indicate the 1-$\sigma$ deviation according to bootstrap resampling of images (and corresponding ground-truth catalogs) in a test set containing 42k images, each $80\times80$ pixels.
The bracketed values indicate the average performance of an inference network across all bins.
For similar plots with performance stratified by signal-to-noise ratio (SNR) rather than magnitude, see Appendix~\ref{sec:snr_results}. 
} \label{fig:eval_multi_field}  

\refstepcounter{panel}\label{fig:precision_multi}
\refstepcounter{panel}\label{fig:recall_multi}
\refstepcounter{panel}\label{fig:class_multi}
\refstepcounter{panel}\label{fig:r_flux_multi}
\end{figure*}

We first compare the performance of our inference networks for light source detection.  
Figures~\ref{fig:precision_multi} and \ref{fig:recall_multi} show the precision (purity) and the recall (completeness) of the detections. The precision is binned by the estimated source r-band magnitude, while the recall (as with all other plots in this section) is binned by the true source r-band magnitude. 
As expected, the detection performance of all three networks increases monotonically for brighter sources. At almost all magnitudes, the PSF-unaware and PSF-aware networks, which are shown with red squares and orange triangles, respectively, outperform the single-field network, which is shown with blue circles.  The exception is in the detection precision for sources in the second-faintest bin, where the networks perform comparably well (within the 1-$\sigma$ bootstrap interval).
Interestingly, the addition of PSF information to evaluation images through parameter-based encoding in the PSF-aware network does little to improve source detection at any magnitude. This implies that to mitigate false detections and detect sources fainter than an r-band magnitude 21.23, the network most benefits from seeing a wider distribution of backgrounds and PSFs; it does not need access to the image-specific background and PSF for robust performance in source detection.

\review{We note that precision and recall remain slightly imperfect even for the brightest bins due to a combination of factors: (1) close pairs of bright sources can blend and be merged or misassigned by the network; (2) our current centroid‐matching uses a strict 1 pixel radius (increasing this to 2–3 pixels would recover additional matches); and (3) as described in Section~\ref{sec:variational-family}, the 4 $\times$ 4 pixel tiling scheme permits at most one detection per tile, so sources straddling tile boundaries are occasionally dropped. We plan to address these limitations in future work.}

\subsubsection{Star/Galaxy Separation} \label{sec:classification_results}

In Figure~\ref{fig:class_multi}, we show the accuracy of the star/galaxy separation of light sources.  If the posterior source-type probability (see Table~\ref{tab:dist_factors}) for an object is greater than 0.5, we classify it as a galaxy; otherwise, we classify it as a star. We further investigate the choice of this classification threshold in Section~\ref{sec:classification_calibration}. Again, at all r-band magnitudes shown, the PSF-unaware and PSF-aware networks outperform the single-field network in accuracy.  

For this task, the PSF-aware network outperforms the PSF-unaware network for r-band magnitudes \review{fainter} than 21, but has comparable performance at other magnitudes. Note that most objects with r-band magnitudes \review{brighter} than $19$ are likely to be stars, whereas those with r-band magnitudes \review{fainter} than $21$ are likely to be galaxies.
This suggests that the explicit incorporation of PSF information may aid in identifying galaxies more than stars.
Interestingly, the single-field network does not exhibit monotonic behavior with source magnitude.  Instead, its classification performance decreases at intermediate magnitudes 19 and 21.5, where we might expect star/galaxy mixing in a magnitude range where the classes are relatively more balanced than at the extremes.  The PSF-unaware and PSF-aware networks mitigate this mixing.

\subsubsection{Flux Measurement} \label{sec:flux_results}
Figure~\ref{fig:r_flux_multi} compares performance for flux prediction, quantified by the mean absolute percentage error (MAPE):
\begin{align}
    \textrm{MAPE} = \frac{1}{n} \sum_{i=1}^{n} \frac{\left|f_\mathrm{true}^{(i)}-f_\mathrm{pred}^{(i)}\right|}{f_\mathrm{true}^{(i)}} \times 100. \label{eq:mape}
\end{align}
\review{We additionally overlay a theoretical lower bound on the flux error, derived from the expected flux uncertainty for isolated sources using real SDSS parameters (gain, background, PSF width, and flux calibration) for run 94, camcol 1, field 12. This bound approximates the minimal achievable MAPE due to photon and background noise.}

We find that both the PSF-unaware and PSF-aware networks outperform the single-field network at all magnitudes. The PSF-unaware network is marginally better than the PSF-aware network for the brightest objects (r-band magnitude $<17.78$), and the PSF-aware network is marginally better for fainter objects (r-band magnitude 21.23--21.75).

All three models are generally better at estimating the flux of brighter objects, with greater error at \review{fainter} magnitudes. The one exception to this is the single-field model in the brightest bin (r-band magnitude $<17.78$). As most objects in this magnitude bin tend to be stars (which look like the PSF), this suggests that seeing a variety of PSFs during training aids in improving flux measurement of these sources in particular.

\subsection{Similar Backgrounds and PSFs}
In Section~\ref{sec:point-estimates}, we evaluated our three inference networks with synthetic data sampled from prior $p_{\text{many}}$. Now, we evaluate all three networks with data sampled from prior $p_{\text{one}}$, which all have the background and PSF of a single SDSS field that is fairly typical in terms of PSF width and background level. These data are sampled from the same distribution as the data used for training the single-field network, but not the other two networks.

\begin{figure*}
\centering
\begin{minipage}{0.49\linewidth}
    \centering
    \includegraphics[width=\linewidth]{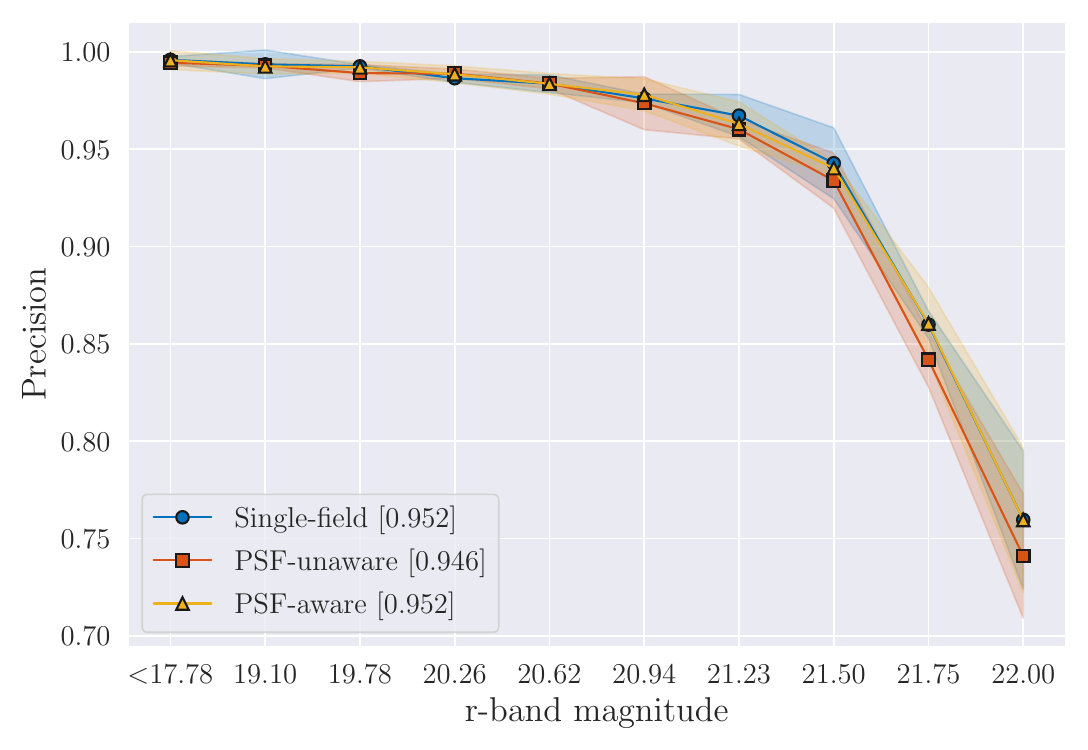}
    \vspace{1ex}
    (a) Precision for light source detection.
\end{minipage}
\hfill
\begin{minipage}{0.49\linewidth}
    \centering
    \includegraphics[width=\linewidth]{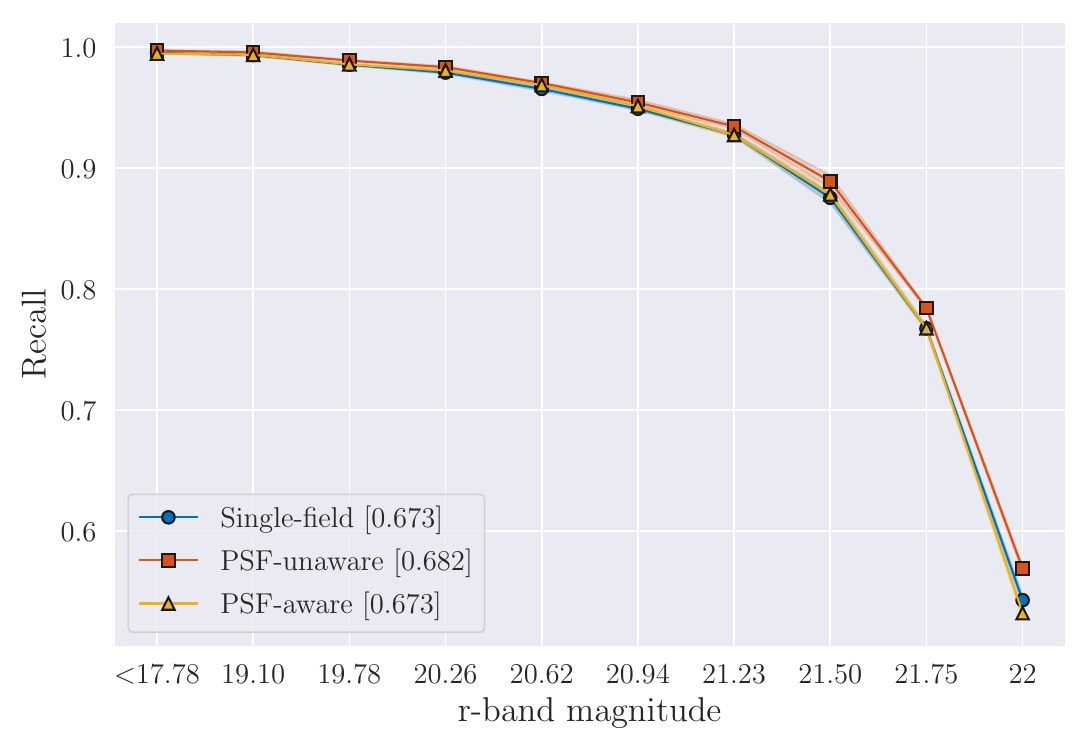}
    \vspace{1ex}
    (b) Recall for light source detection.
\end{minipage}
\vspace{2ex}
\begin{minipage}{0.49\linewidth}
    \centering
    \includegraphics[width=\linewidth]{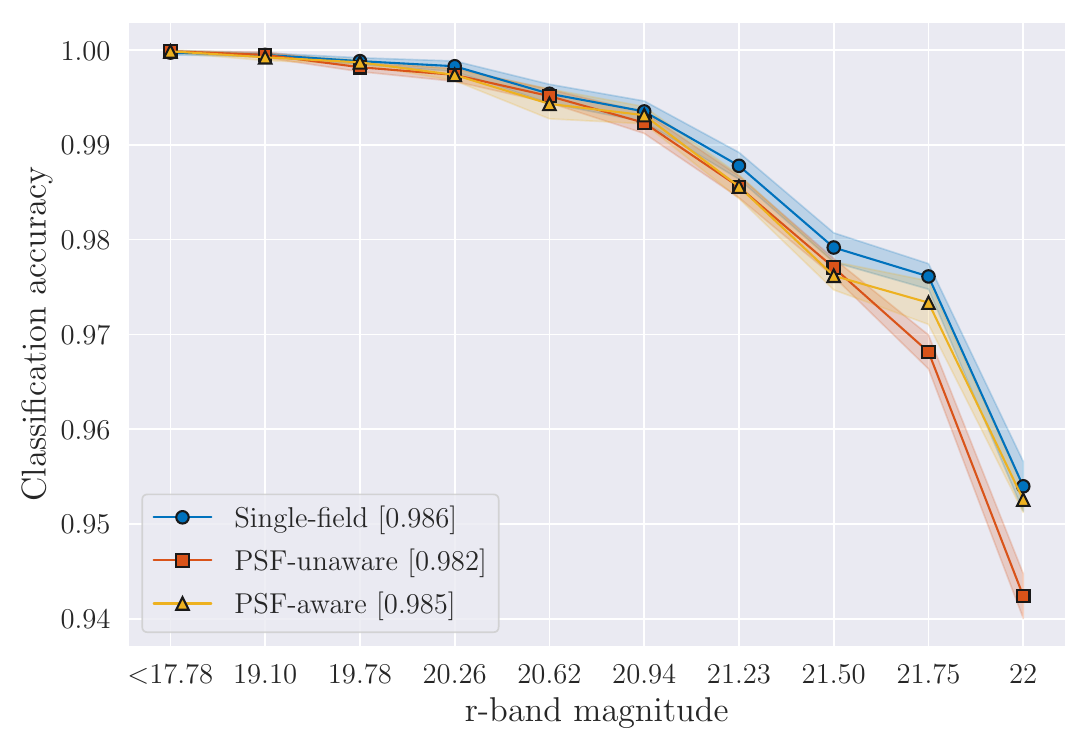}
    \vspace{1ex}
    (c) Accuracy for star/galaxy separation.
\end{minipage}
\hfill
\begin{minipage}{0.49\linewidth}
    \centering
    \includegraphics[width=\linewidth]{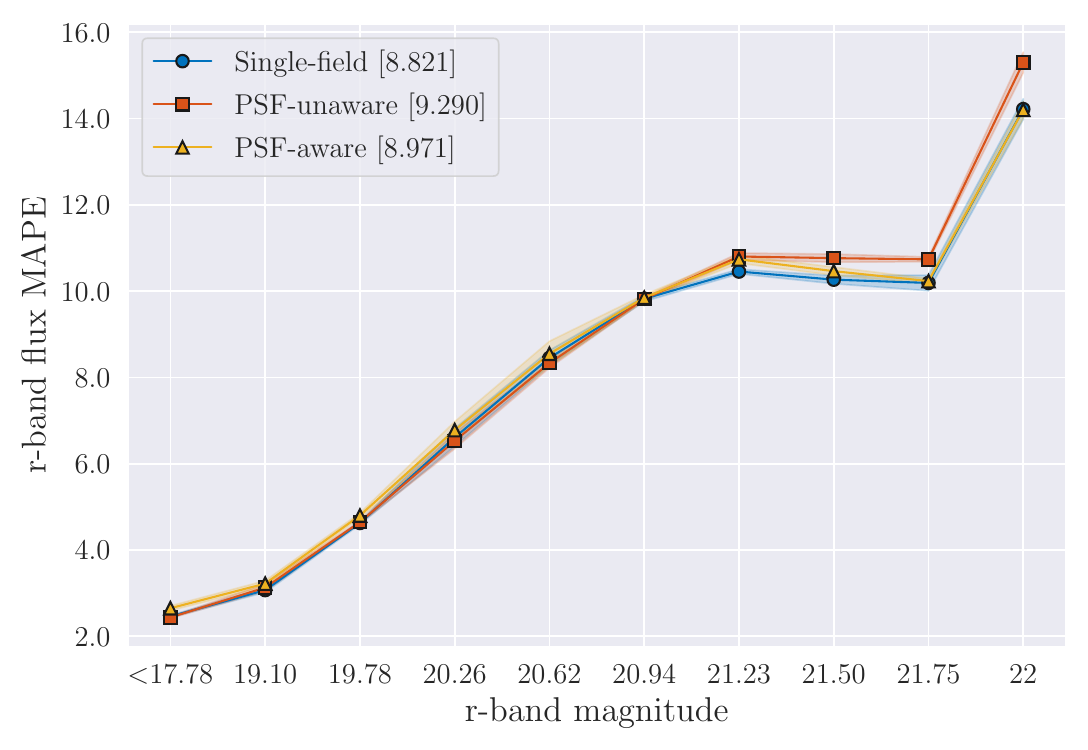}
    \vspace{1ex}
    (d) Mean absolute percentage error for flux estimation.
\end{minipage}
\vspace{2ex}
\caption{\textbf{Performance for held-out data with a fixed background and PSF.}
As in Figure~\ref{fig:eval_multi_field}, solid lines indicate the mean performance in each r-band magnitude bin, the widths of the bins are selected so that each bin contains an equal number of objects, the shaded regions indicate the 1-$\sigma$ deviation according to bootstrap resampling, and the bracketed values indicate the average performance of an inference network across all bins. In contrast to the setting of Figure~\ref{fig:eval_multi_field}, here all networks perform similarly.}
\label{fig:eval_single_field}
\end{figure*}

Figure~\ref{fig:eval_single_field} shows the same metrics for detection, star/galaxy separation, and flux measurement as in Section~\ref{sec:point-estimates}, but now evaluated with this new data distribution.  In all tasks, the PSF-unaware and PSF-aware networks are competitive with the single-field network, despite being evaluated on \review{a test dataset sampled from a different distribution than the training set}.  In fact, all three models are nearly identical at \review{bright} ($<21$) magnitudes for all three metrics. The PSF-unaware network has slightly poorer precision, classification accuracy, and flux MAPE at fainter magnitudes than the PSF-aware and single-field models, which are consistently within 1-$\sigma$ of each other.


In most cases, the performance of the single-field network matches or exceeds the performance of the PSF-aware and PSF-unaware networks. However, the PSF-unaware network has slightly higher recall for faint objects ($\geq21.5$ r-band magnitude). A potential explanation for this is that detecting faint objects in bright backgrounds is harder than detecting faint objects in low-intensity backgrounds. By training the generalized networks with a variety of backgrounds, the network improves its overall detection ability, not just for performing inference with a brighter background.

\subsection{Calibration of posterior approximations}
Here, we assess the calibration of the posterior approximations produced by each inference network. 

\subsubsection{Detection}
\label{sec:detection_calibration}

\begin{figure*}
    \centering
    \includegraphics[width=\linewidth]{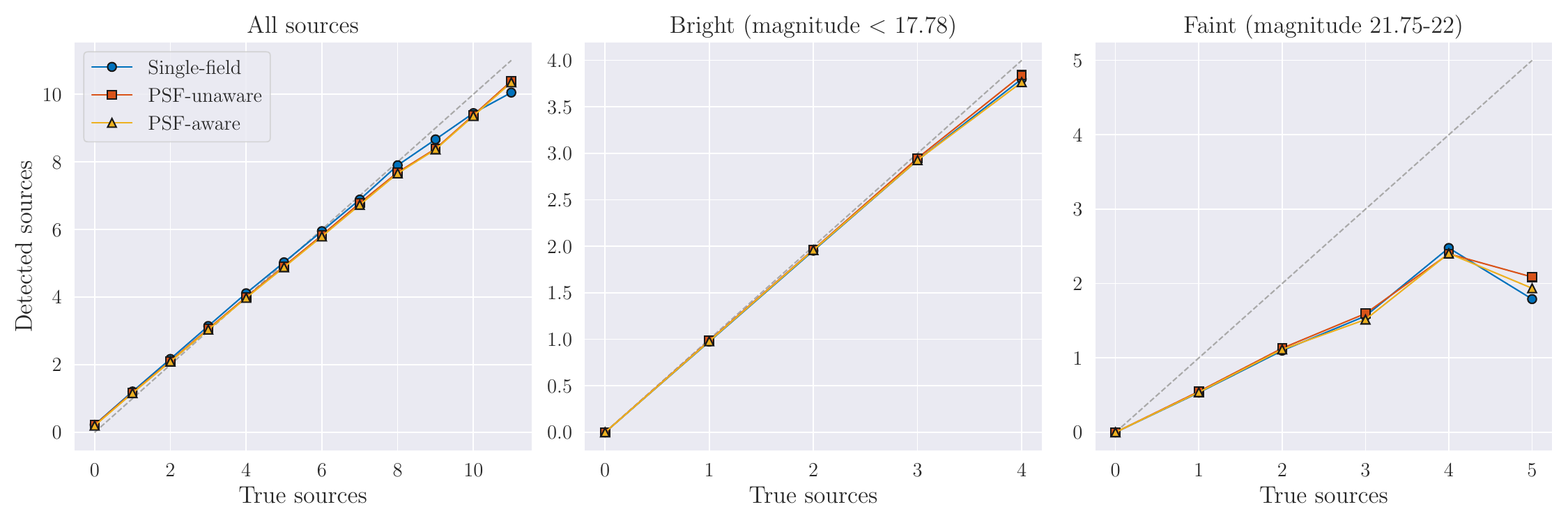}
    \caption{\textbf{The expectation number of detections vs. the true number of sources.} Left: all magnitudes. Center: the brightest sources ($<$17.78 magnitude). Right: the faintest sources (21.75--22 magnitude). Each point represents the average over all images with a particular number of sources. The diagonal dotted gray line represents perfect performance, where the number of sources always equals the number of detections. All three networks perform equally well on bright sources and are underconfident on faint sources.}
    \label{fig:true_vs_expected_sources}
\end{figure*}

Figure~\ref{fig:true_vs_expected_sources} shows calibration for source detection.  The figure panels show the expected number of sources in an 80 $\times$ 80 pixel image, given by summing the probabilities of detections, as a function of the actual number of sources in that image.  From left to right, we show this for all sources, the top 10th percentile of the brightest sources ($<17.78$ magnitude), and the bottom 10th percentile of the faintest sources ($21.75$--$22$ magnitude). The dashed gray line traces the behavior of a perfectly calibrated network, which would detect the exact number of sources.

Overall, all three networks are fairly well-calibrated; that is, the expected value of detected sources is almost equal to the number of actual sources, with some underconfidence when there are many sources in an image ($>8$).

When restricting the analysis to sources brighter than $17.78$ magnitude, all three networks are nearly perfectly calibrated until there are four bright sources in the image.  This result is consistent with the intuition that the brightest objects are the easiest to detect. There is some under-confidence in cases in which many bright sources appear in an image, likely due to ambiguity when there are many bright sources near one another that begin to visually blend. 

When restricting the analysis to the faintest sources in our dataset (with r-band magnitude between 21.75 and 22), all three networks are underconfident in their predictions, systematically predicting fewer faint objects than truth. 
This is also consistent with the intuition that faint objects are the most difficult to detect. There is a minimal difference between the calibration of the PSF-unaware and PSF-aware networks.  This is consistent with the detection metrics in Figures~\ref{fig:precision_multi} and \ref{fig:recall_multi}, where explicit PSF information also led to limited improvement in overall detection performance.

\subsubsection{Star/Galaxy Separation}  \label{sec:classification_calibration}
\begin{figure*}
    \centering
    \includegraphics[width=\linewidth]{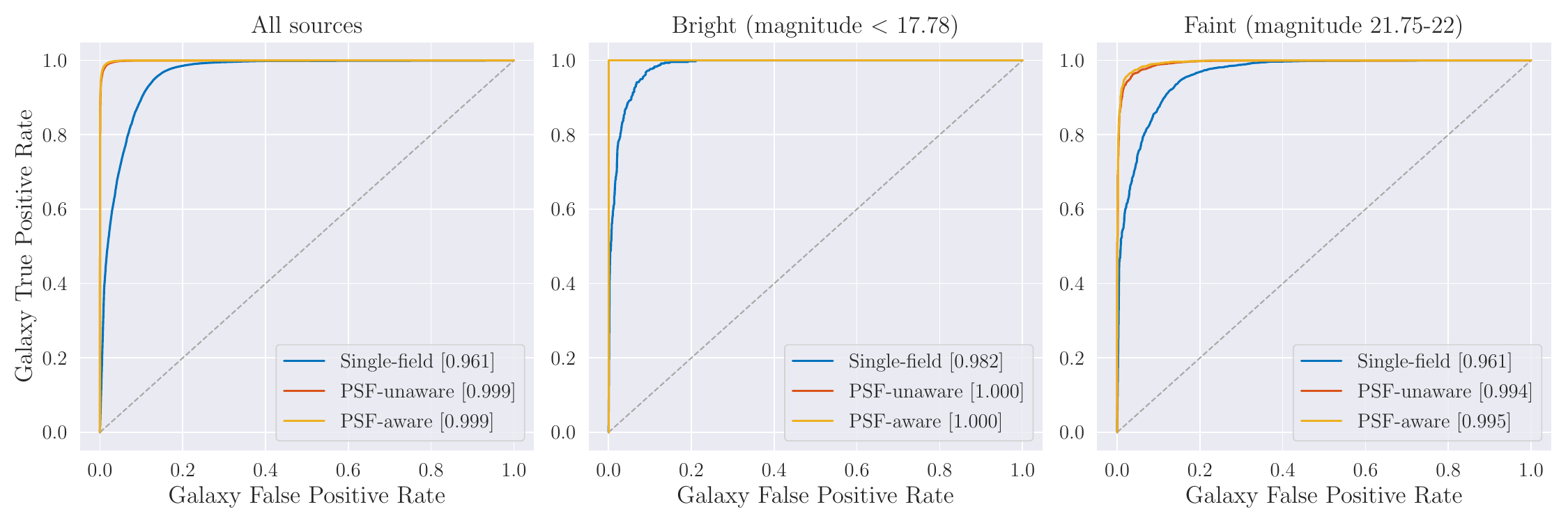}
    \caption{\textbf{Receiver operating characteristic (ROC) curve for star/galaxy separation.} The true positive rate for galaxies (sensitivity) is plotted against the false positive rate (1 - specificity) for different classification thresholds. Left: all magnitudes. Center: the brightest sources (magnitude $<$17.78). Right: the faintest sources (magnitude 21.75--22). The dotted gray line represents the performance of a random classifier. The area under the curve (AUC) for each inference network is provided in brackets. }
    \label{fig:classification_roc}
\end{figure*}

Previously, in Section~\ref{sec:classification_results}, we assessed the accuracy of our star/galaxy separation for predictions determined by applying a single decision threshold (0.5) to the inferred source-type probability. Here, we assess the performance of the networks as the decision threshold is varied. Figure~\ref{fig:classification_roc} shows the receiver operating characteristic (ROC) curve, which illustrates the discriminative performance of a binary classifier as the decision threshold increases. The ROC curve plots the true positive rate (TPR) against the false positive rate (FPR). The area under the ROC curve (AUC) quantifies the overall performance of the classifier, with values closer to 1.0 corresponding to better performance.

We find that the PSF-unaware and PSF-aware models significantly outperform the single-field network at all magnitudes, although the AUC for all three models is respectable. At bright magnitudes, the PSF-unaware and PSF-aware models achieve near-perfect discrimination, consistent with Figure~\ref{fig:class_multi}. At faint magnitudes, the difference between models is more pronounced, with the PSF-aware network showing a slight improvement over the PSF-unaware network.

\subsubsection{Flux Measurement}
\begin{figure*}
    \centering
    \includegraphics[width=\linewidth]{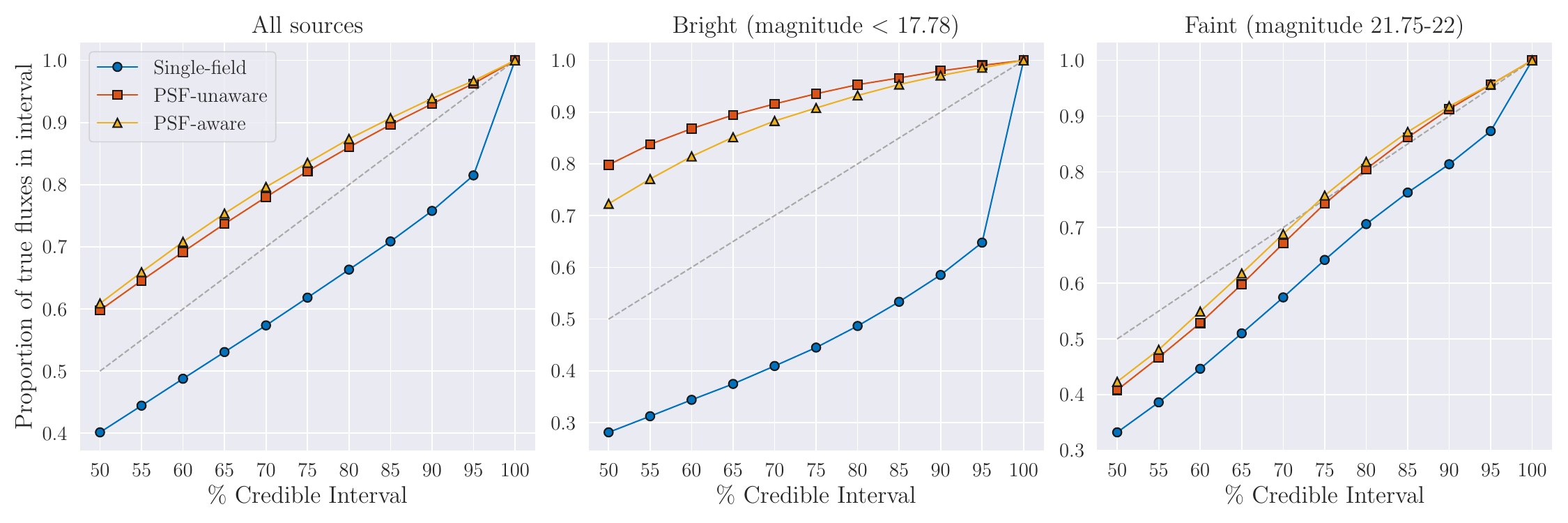}
    \caption{\textbf{Proportion of sources whose flux falls within various credible intervals.} Left: all magnitudes. Center: the brightest sources ($<$17.78 magnitude). Right: the faintest sources (21.75--22 magnitude). The dotted gray line represents perfect performance, where the credible interval contains exactly that percentage of sources.
    }
    \label{fig:flux_in_interval}
\end{figure*}

We now examine how well calibrated our networks are when estimating the flux of each source by comparing the fraction of true flux values contained in the credible interval of the posterior output by our networks.  Credible intervals are the Bayesian analogue of confidence intervals; they describe the probability that the true flux should fall within that interval. For example, for a perfectly calibrated network, a 95\% credible interval contains the true flux 95\% of the time.

Figure~\ref{fig:flux_in_interval} shows the proportion of sources whose flux falls in various credible intervals of the inferred posterior.  The panels show an identical data subselection as in Figures~\ref{fig:true_vs_expected_sources} and \ref{fig:classification_roc}.

The single-field network is underdispersed for all magnitude ranges, meaning that the variance predicted by the network is lower than the observed variance.  An underdispersed network produces overly narrow credible intervals, which would contain the true flux less often than expected.  For example, when evaluated on all sources, the 60\% credible intervals of the single-field network contain only 48\% of the true flux values.
\review{
The generalist networks (PSF-aware and PSF-unaware) are somewhat better calibrated than the specialist network (single-field), but the degree to which the former are better is less than the degree to which all networks are miscalibrated. This shared misspecification can happen for a variety of reasons, but in variational inference, the leading reason is restrictions on the form of the variational distribution, which in this case approximates the flux posterior as log normal. This source of miscalibration may be mitigated by a more flexible variational distribution, such as a mixture of normal distribution, but actually doing so is beyond the scope of this work, which is focused on comparing inference network training regimes.
}

On the other hand, for bright light sources, both the PSF-unaware and PSF-aware networks produce overly wide credible intervals, which contain the true flux more often than expected. For example, when evaluated on all sources, the 60\% credible intervals for the PSF-aware network contain the true flux about 70\% of the time.  For faint magnitudes, the credible intervals between $50\%$ and $75\%$ miss the true flux no more than $10\%$ of the time, and the credible intervals targeting more than $75\%$ coverage are well calibrated, where the proportion of true flux values contained in the predicted credible interval closely matches the nominal coverage probability. For instance, both networks' 95\% credible intervals contain the true flux approximately 96\% of the time, indicating that the uncertainty is accurately quantified and the intervals are neither too conservative nor too narrow.

We find that the PSF-aware network is more poorly calibrated than the PSF-unaware network for all sources, but better calibrated for the brightest and faintest sources. Furthermore, all three networks are best calibrated for faint sources and more poorly calibrated for bright sources. This suggests that although predicting the center of the distribution is easier for the brightest objects (see Figure~\ref{fig:r_flux_multi}), uncertainty estimation is more error-prone at these magnitudes.


\section{Discussion}  \label{sec:discussion}
In this work, we applied neural posterior estimation to images with spatially varying backgrounds and PSFs.
We considered two inference network architectures: one that accepted estimated PSF parameters as input and one that did not.
We trained inference networks based on both architectures with varied backgrounds and PSFs from many SDSS fields.
We compared these networks with a baseline network, which did not accept estimated PSF parameters as input and was trained using only the background and PSF from a single SDSS field.

\subsubsection*{Summary of our findings}
Training inference networks on simulated images with varied backgrounds and PSFs aids in detection performance by increasing robustness across fields. 
Training with varied PSFs and backgrounds also improves calibration for star/galaxy separation and flux measurement.
In all tasks, inference networks trained with a variety of PSFs and backgrounds performed no worse than an inference network trained with a single background and PSF, even when evaluated on images generated with this specific background and PSF (see Figure~\ref{fig:eval_single_field}).

Providing an inference network with PSF information (e.g., through dedicated input channels) can improve performance in star/galaxy separation and flux measurement. The most noticeable improvements were in star/galaxy separation for objects of intermediate brightness and in flux measurement for the brightest objects (see Figures~\ref{fig:precision_multi}, \ref{fig:recall_multi}, and \ref{fig:flux_in_interval}). These results for light source detection and star/galaxy separation, obtained with explicit PSF information,  are comparable to those reported for other image analysis methods such as \citet{shi2022photometry, sevilla2018star, clarke2020identifying}. Specifically, our detection precision exceeds 0.95 for r-band magnitudes brighter than 21.5, and our detection recall exceeds 0.9 for r-band magnitudes brighter than 21.2.  Our star/galaxy separation accuracy exceeds 0.96 for r-band magnitudes brighter than 22.
However, little improvement in detection performance is gained by providing explicit information about the background and PSF to the inference network (see Figures~\ref{fig:precision_multi} and \ref{fig:recall_multi}).

Although the single-field network was often outperformed by the other networks, we were surprised that it performed as well as it did in our experiments, particularly for detection, given that it was not trained with a variety of PSFs and background intensities (see Figures \ref{fig:precision_multi}, \ref{fig:recall_multi}, and \ref{fig:true_vs_expected_sources}). 
\review{
The conventional wisdom is that neural networks are extremely brittle when faced with inputs outside of the support of the distribution of their training data, and indeed, we observed some evidence of this in that the single-field network was always outperformed in these settings.
However, this conventional wisdom lacks nuance about how far beyond the support of the training distribution a particular input is. For our particular network architecture and the degree of PSF and background variation in observed in SDSS, this conventional wisdom, though broadly correct, may overstate the concern.
}

\subsubsection*{Limitations and future work}
The scope of this work is the application of NPE to a model with random backgrounds and PSFs for the three specific tasks presented.
We have several extensions of our method and our analysis of it in mind for future work. 

\textit{Enhanced PSF encoding.}
In this work, we used the six-parameter PSF encoding provided by the SDSS pipeline, which represents the PSF as \review{isotropic}.
However, the true PSF is unlikely to be exactly \review{isotropic} or otherwise perfectly characterized by six parameters.
While the six-parameter PSF encoding may be adequate in some applications, a more sophisticated PSF encoding could be obtained using an auxiliary autoencoder, as in \citet{jia2020point}.  By allowing an autoencoder to learn a compressed representation of the PSF, we allow for greater expressivity than a simple parametric model. 

\textit{Application to deeper surveys.}
We conducted our experiments with synthetic data designed to mirror SDSS data.
Mirroring the data of a well-understood survey simplified the generation of realistic synthetic data.
For instance, we could fit our prior distribution to existing catalogs without needing to extrapolate to what sources would be detectable in images of greater depth.

However, the SDSS observes a lower density of detectable light sources than upcoming surveys such as the LSST.
Thus, the precise extent to which our results are relevant to upcoming surveys is somewhat ambiguous.
Although a priori we do not see why the conclusions of our present work, which compares methods of incorporating spatially varying covariates, would be specific to a particular source density, it is to some extent an empirical question.

However, if our primary objective were to evaluate NPE for cataloging in general, rather than NPE for cataloging specifically in the context of spatially varying covariates (that is, background and PSF), then the relevance of our results to upcoming surveys would depend more on the source density used in our experiments. In ongoing work, we are applying NPE to the LSST-like DC2 simulated dataset \citep{abolfathi2021lsst}, so far with promising preliminary results: the value of NPE appears to increase with source density, as the greater ambiguity stemming from the presence of more blended sources favors probabilistic approaches.

\textit{Quantifying deblending performance.}
In Sections~\ref{sec:detection_results} and \ref{sec:detection_calibration}, we examine the extent to which our method infers the correct number of light sources.  This is a particularly challenging problem in crowded fields where light sources overlap and blend.
In ongoing work, we are further investigating how detection performance varies with blendedness, in relation to established deblending methods such as Source Extractor \citep{bertin1996sextractor} and Scarlet \citep{melchior2018scarlet}.
Because this work focuses on spatially varying covariates (background and PSF), we do not explicitly consider the degree of blendedness in our metrics for detection or for other tasks.

\review{
\textit{Mitigating model misspecification.}
Imperfect simulators pose two distinct problems for neural posterior estimation (NPE). First, like most inference methods, simulation mismatch yields a posterior that is inaccurate proportional to the misspecification. Second---and specific to NPE---the inference network may be presented at inference time with out‑of‑distribution data.
}

\review{
In this work, we sidestep both issues by training and evaluating exclusively on synthetic data from the same simulator, to address the challenges of spatially varying covariates in isolation of issues of model misspecification.
In the application of NPE to real astronomical images, some degree of model misspecification is inevitable.
Previous work applying NPE to crowded starfields, using real SDSS images, did not require any special techniques to overcome model misspecification beyond carefully tuning the simulator~\citep{liu2023variational}. However, a variety of techniques have been developed to mitigate model misspecification in NPE.
For example, \citet{ward2022robust} shows that naïve NPE can produce unreliable posterior approximations under misspecification and proposes to model a simulator–target error channel that ``denoises'' observations and recovers credible posterior estimates. 
\citet{huang2023learning} shows that summary‑statistic regularization in NPE suppresses simulator‑sensitive features and stabilizes both embeddings and posteriors under mismatch.
}

\review{
Finally, we note that inputting imperfect pipeline estimate of PSF into our PSF-aware inference networks is largely a separate issue from model misspecification, and one that poses little challenge for our methods. During training, an inference network naturally learns to adjust or ignore misleading inputs, without requiring any explicit domain adaptation.
}

\textit{Measuring galaxy shapes.}
To assess the success of our integration of spatially varying covariates (background and PSF) with NPE, we considered three tasks: detection, star/galaxy separation, and flux measurement. 
We are intrigued by the possibility of using NPE for galaxy-shape measurement in the context of properly modeling PSF. 
A proper accounting of the PSF should be particularly helpful in estimating the half-light radius.
However, we leave this to follow-up work, as our current three tasks seem adequate to demonstrate NPE with spatially variable covariates.
One challenge in benchmarking galaxy shape measurement is in defining scientifically relevant performance metrics. 
If the ultimate use of these galaxy shape measurements is to estimate shear-shear correlation for weak lensing analyses, as is often the case, then average ellipticity error may not be a meaningful measure of error.

\subsubsection*{Conclusion}
NPE is an effective means of analyzing astronomical images with spatially varying covariates, such as the background and the PSF.
An inference network trained with a variety of backgrounds and PSFs can efficiently perform probabilistic light source detection, star/galaxy separation, and flux measurement.

\section*{Author Contributions}
AP and JR extended the BLISS software to model variable PSFs and backgrounds.
AP conducted the numerical experiments and generated the figures.
CA and JR conceived of the project and supervised the work.
AP, TZ, CA, and JR wrote the text of this manuscript.

\section*{Acknowledgments}
This article builds on the work of those who have contributed code to the BLISS project.
We are particularly grateful to those who have made foundational contributions to the BLISS project, including Runjing Liu, Ismael Mendoza, Derek Hansen, Zhe Zhao, Ziteng Pang, Yash Patel, Xinyue Li, Sawan Patel, and Zhixiang Teoh.

This paper has undergone internal review in the LSST Dark Energy Science Collaboration. The internal reviewers were Bob Armstrong and Shuang Liang. 

This material is based upon work supported by the U.S. Department of Energy, Office of Science, Office of High-Energy Physics under Award Number DE-SC0023714. TZ is supported by Schmidt Sciences.

\input{standard_ack}

\section*{Data Availability}
Our software is available in a public GitHub repository at \url{https://github.com/prob-ml/bliss} \review{and a static version is 
 posted on Zenodo \citep{patel_et_al_2025_15779060}}.
The specific code for reproducing the figures and tables in this article is within the \texttt{case\_studies/psf\_variation} directory.

\bibliography{references}{}
\bibliographystyle{aasjournal}



\appendix

\counterwithin{figure}{section}
\counterwithin{table}{section}
\renewcommand{\thefigure}{\thesection.\arabic{figure}}

\section{Performance stratified by Signal-to-Noise Ratio}  \label{sec:snr_results}

In this appendix, we repeat our evaluation from Section~\ref{sec:point-estimates} but report performance stratified by the signal-to-noise ratios (SNRs) of objects rather than the magnitudes of the objects. The SNR of an object depends on the flux within the object's footprint, noise from the background, and Poisson noise of the object. We determine the footprint by masking out all pixels whose fluxes did not appreciably change with the addition of a new object. Then, following \citet{Jones_2024}, we compute the SNR as follows:
\begin{align*}
    \mathrm{SNR} = \frac{C}{\sqrt{C + B + P}}, \label{eq:snr}
\end{align*}
where $C$ is the source counts in the footprint, $B$ is the background counts in the footprint, and $P$ is the Poisson noise in the footprint.

\begin{figure*}[h]
\centering

\begin{minipage}{0.49\linewidth}
    \centering
    \includegraphics[width=\linewidth]{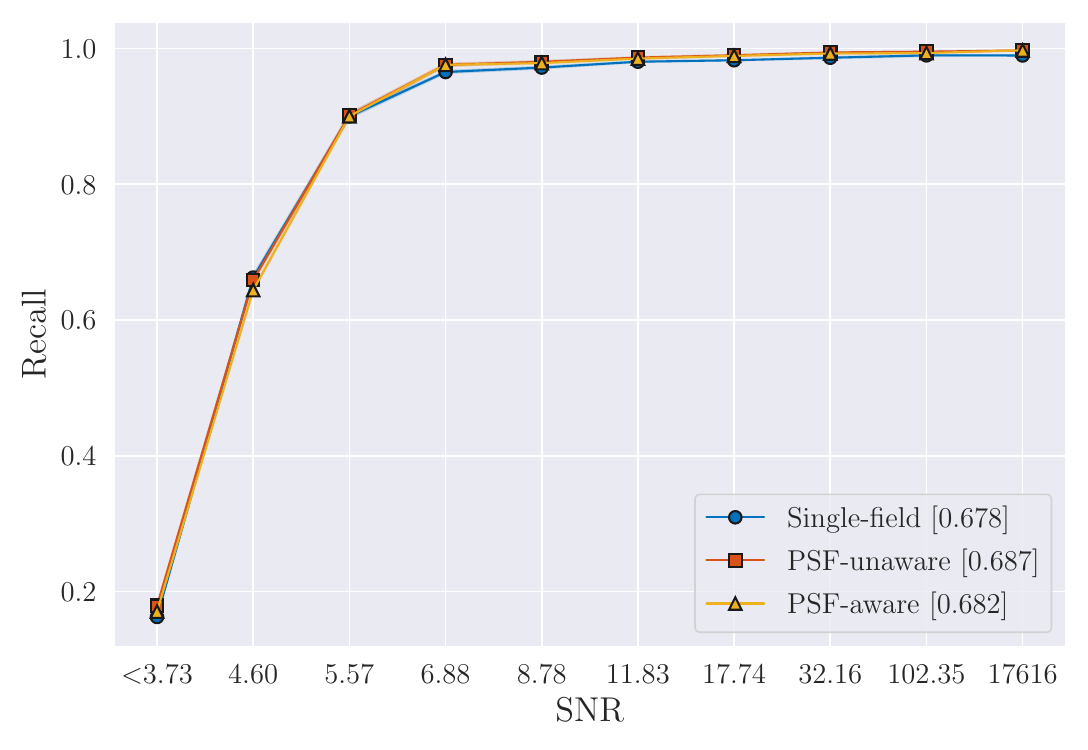}
    \vspace{1ex}
    
    (a) Recall for light source detection.
    \label{fig:recall_snr}
\end{minipage}
\hfill
\begin{minipage}{0.49\linewidth}
    \centering
    \includegraphics[width=\linewidth]{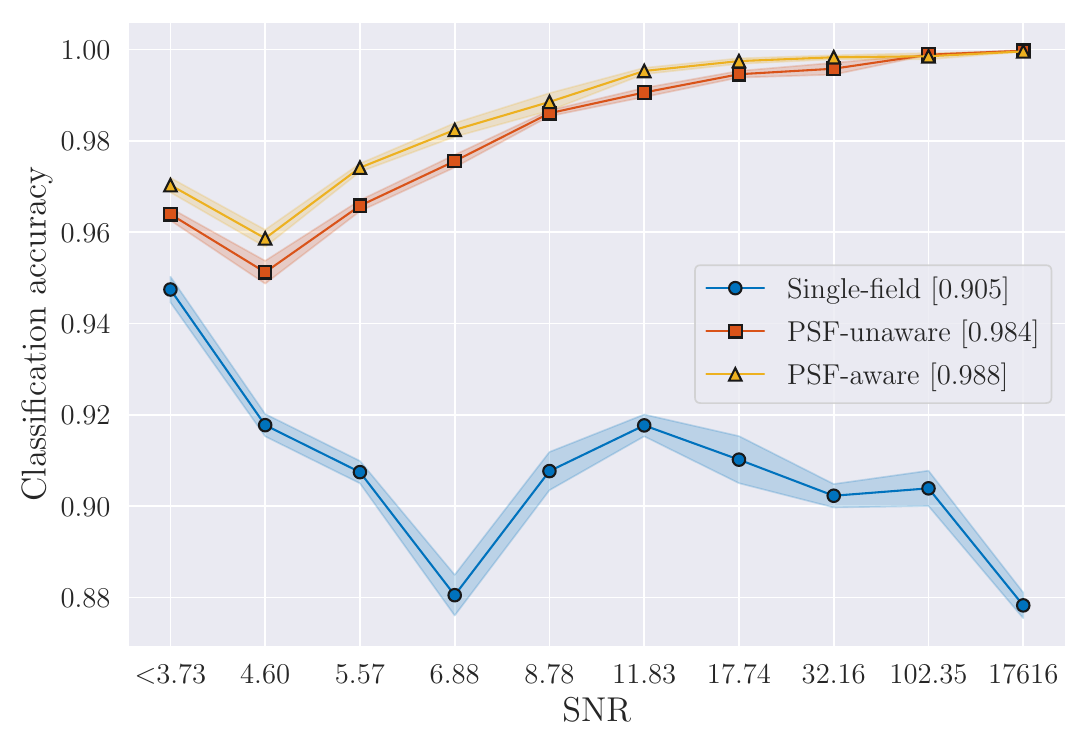}
    \vspace{1ex}
    
    (b) Accuracy for star/galaxy separation.
    \label{fig:class_multi_snr}
\end{minipage}

\vspace{2ex}

\begin{minipage}{0.49\linewidth}
    \centering
    \includegraphics[width=\linewidth]{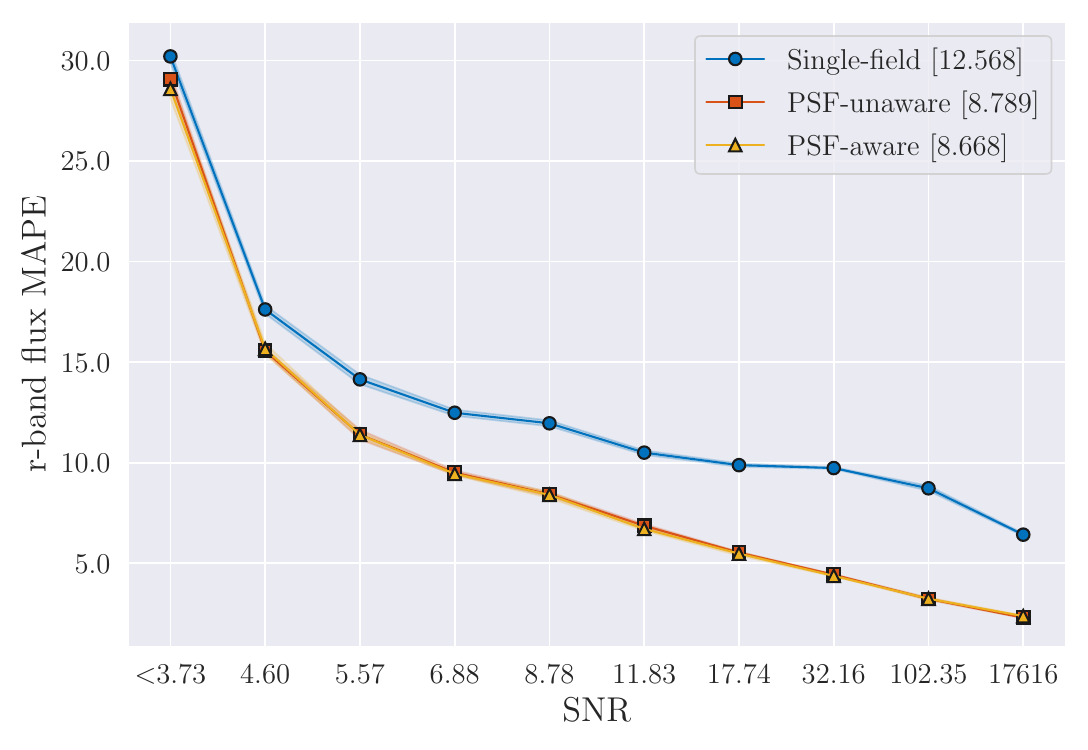}
    \vspace{1ex}
    
    (c) Mean absolute percentage error for flux estimation.
    \label{fig:r_flux_multi_snr}
\end{minipage}

\vspace{2ex}

\caption{\textbf{Performance for held-out data with varied backgrounds and PSFs.}
This figure displays the data from Figure~\ref{fig:eval_multi_field} with performance stratified by signal-to-noise (SNR) rather than object magnitude.}
\label{fig:eval_multi_field_snr}
\end{figure*}

Figure \ref{fig:eval_multi_field_snr} is analogous to Figure \ref{fig:eval_multi_field} in the main text, with performance stratified by signal-to-noise (SNR) rather than object magnitude. Precision for light source detection is omitted from Figure \ref{fig:eval_multi_field_snr} because it is unclear how to compute the SNR of unmatched detections.

Like Figure \ref{fig:eval_multi_field}, Figure \ref{fig:eval_multi_field_snr} shows 1) nearly monotonic performance of the PSF-unaware and PSF-aware networks, 2) the single-field network often performing worse than the PSF-unaware and PSF-aware networks without ever performing better, and 3) the PSF-aware network sometimes outperforming the PSF-unaware network without ever performing worse.

\begin{figure}[h]
    \centering
    \includegraphics[width=\linewidth]{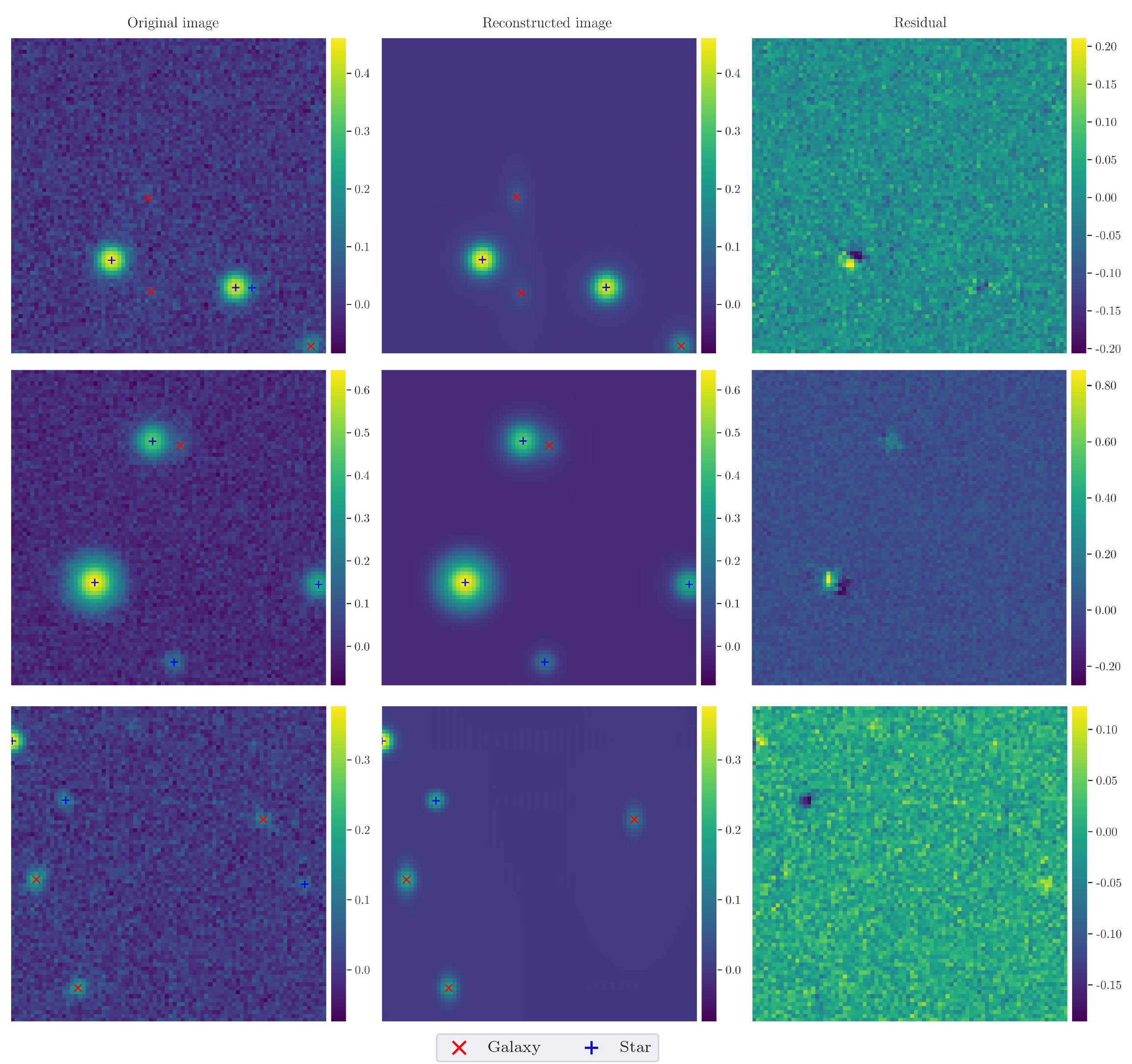}
    \caption{Three examples of cataloging with the ``PSF-aware'' inference network, one per row. The left column shows input images, which are held-out and semi-synthetic, along with the true positions and types (star vs galaxy) of the imaged astronomical objects. The center column show our predictions of the positions and types of astronomical objects in these images (that is, mode of variational distribution), as well as model reconstructions of the images. The right column shows the residual for each pixel.}
    \label{fig:reconstruction}
\end{figure}

\clearpage
\review{
\section{Image Reconstruction} \label{sec:reconstruction}
In Figure~\ref{fig:reconstruction}, we show samples of astronomical images from a held-out semi-synthetic dataset and the catalogs of them produced by our inference network.
}

\end{document}

%% file: standard_ack.tex
The DESC acknowledges ongoing support from the Institut National de 
Physique Nucl\'eaire et de Physique des Particules in France; the 
Science \& Technology Facilities Council in the United Kingdom; and the
Department of Energy, the National Science Foundation, and the LSST 
Corporation in the United States.  DESC uses resources of the IN2P3 
Computing Center (CC-IN2P3--Lyon/Villeurbanne - France) funded by the 
Centre National de la Recherche Scientifique; the National Energy 
Research Scientific Computing Center, a DOE Office of Science User 
Facility supported by the Office of Science of the U.S.\ Department of
Energy under Contract No.\ DE-AC02-05CH11231; STFC DiRAC HPC Facilities, 
funded by UK BEIS National E-infrastructure capital grants; and the UK 
particle physics grid, supported by the GridPP Collaboration.  This 
work was performed in part under DOE Contract DE-AC02-76SF00515.